\documentclass{emulateapj}
\usepackage{natbib}
\usepackage{apjfonts}
\begin{document}

\title{The Supermassive Black Hole in M84 Revisited\footnotemark[1]}

\author{Jonelle L. Walsh\footnotemark[2], Aaron
J. Barth\footnotemark[2], Marc Sarzi\footnotemark[3]}

\footnotetext[1]{Based on observations made with the NASA/ESA
\emph{Hubble Space Telescope}, obtained from the Data Archive at the
Space Telescope Science Institute, which is operated by the
Association of Universities for Research in Astronomy, Inc., under
NASA contract NAS 5-26555. These observations are associated with
program GO-7124 and GO-6094.}

\footnotetext[2]{Department of Physics and Astronomy, University of
California at Irvine, 4129 Frederick Reines Hall, Irvine, CA
92697-4574; jlwalsh@uci.edu, barth@uci.edu}

\footnotetext[3]{Centre for Astrophysics Research, University of
Hertfordshire, AL10 9AB Hatfield, UK; sarzi@star.herts.ac.uk}

\begin{abstract}

The mass of the central black hole in the giant elliptical galaxy M84
has previously been measured by two groups using the same observations
of emission-line gas with the Space Telescope Imaging Spectrograph
(STIS) on the \emph{Hubble Space Telescope}, giving strongly
discrepant results: \citet{Bower_1998} found $M_\mathrm{BH} =
(1.5^{+1.1}_{-0.6}) \times 10^{9}\ M_\odot$, while
\citet{Maciejewski_Binney_2001} estimated $M_\mathrm{BH} = 4 \times
10^8\ M_\odot$. In order to resolve this discrepancy, we have
performed new measurements of the gas kinematics in M84 from the same
archival data, and carried out comprehensive gas-dynamical modeling
for the emission-line disk within $\sim 70$ pc from the nucleus. In
comparison with the two previous studies of M84, our analysis includes
a more complete treatment of the propagation of emission-line profiles
through the telescope and STIS optics, as well as inclusion of the
effects of an intrinsic velocity dispersion in the emission-line
disk. We find that an intrinsic velocity dispersion is needed in order
to match the observed line widths, and we calculate gas-dynamical
models both with and without a correction for asymmetric
drift. Including the effect of asymmetric drift improves the model fit
to the observed velocity field. Our best-fitting model with asymmetric
drift gives $M_\mathrm{BH} = (8.5^{+0.9}_{-0.8}) \times 10^8\ M_\odot$
(68\% confidence). This is a factor of $\sim 2$ smaller than the mass
often adopted in studies of the $M_\mathrm{BH} - \sigma_\star$ and
$M_\mathrm{BH} - L$ relationships. Our result provides a firmer basis
for the inclusion of M84 in the correlations between black hole mass
and host galaxy properties.

\end{abstract}


\section{Introduction}
\label{sec:intro}

Located in the Virgo cluster, M84 (NGC 4374) is an elliptical galaxy
with a radio-loud active galactic nucleus (AGN). According to
\cite{Ho_1997}, the AGN is optically classified as a Type 2
low-ionization nuclear emission-line region (LINER). M84 is known to
contain a circumnuclear gas disk \citep{Bower_1997}, providing the
opportunity to measure the mass of the central black hole if the gas
participates in circular rotation in a thin disk-like structure.
Following the installation of the Space Telescope Imaging Spectrograph
(STIS) on the \emph{Hubble Space Telescope} (\emph{HST}), M84 was the
first target for a gas-dynamical measurement of its black hole mass
\citep{Bower_1998}.  Gas-dynamical measurements had previously been
made for several objects using data from the Faint Object Spectrograph
(FOS) and Faint Object Camera (FOC) \citep{Harms_1994, Ferrarese_1996,
Macchetto_1997, vanderMarel_vandenBosch_1998, Ferrarese_1999}, but
STIS provided a dramatic improvement in data quality for
spatially-resolved emission-line spectroscopy.

\cite{Bower_1998} showed that the M84 gas kinematics could be modeled
as a disk in circular rotation. However, near the M84 nucleus, the
STIS spectra are very complicated with severely blended,
double-peaked, and asymmetric H$\alpha$ and [\ion{N}{2}] line
profiles. \cite{Bower_1998} interpreted the complex nature of the
emission lines as arising from two kinematically distinct gas
components. One high velocity gas component was thought to be part of
the nuclear gas disk rotating about the black hole, while the other
low velocity component was believed to be an extension of a
filamentary structure seen on larger scales. \cite{Bower_1998}
therefore included only the high velocity component in their modeling,
and determined a black hole mass of $(1.5^{+1.1}_{-0.6}) \times
10^{9}\ M_\odot$.

An alternative explanation for the complex line profiles seen near the
M84 nucleus was suggested by \cite{Maciejewski_Binney_2001}. They
argued that a rapidly rotating disk could give rise to the complex
line profiles and that the complicated spectra are due to a slit width
that is larger than the core of the telescope point-spread function
(PSF). The location at which light enters the slit affects the
inferred velocity, creating an instrumental velocity gradient along
the dispersion direction. In the case of M84,
\cite{Maciejewski_Binney_2001} identified a ``caustic'' that occurs
where the instrumental velocity offset balances the change in velocity
due to the gas rotating about the black hole. They used the location
of the caustic to estimate a black hole mass of $4 \times 10^8\
M_\odot$.  Similarly, \cite{Barth_2001b} attempted to model the
complex line profiles in M84 and found that the double-peaked and
asymmetric profiles could be qualitatively reproduced with a single,
rapidly rotating disk component observed through the STIS
0\farcs2-wide slit, however they were unable to determine a black hole
mass because a direct fit of the modeled line profiles to the data was
not well constrained.

M84 has a stellar velocity dispersion of $296\pm14$ km s$^{-1}$, and
the black hole in M84 lies near the high-mass end of the correlations
between black hole mass and host-galaxy properties, such as those with
the bulge stellar velocity dispersion ($M_\mathrm{BH} - \sigma_\star$;
\citealt{Ferrarese_2000,Gebhardt_2000,Tremaine_2002, Gultekin_2009})
and bulge luminosity ($M_\mathrm{BH} - L$;
\citealt{Kormendy_Gebhardt_2001, Gultekin_2009}). These relationships
imply that the growth of black holes and bulges are intimately linked,
and the correlations have crucial implications for galaxy formation
and evolution. Properly determining black hole masses at the high-mass
end of the relationships is particularly important because those
measurements will have a substantial effect on the slope and scatter
of the relationships, which in turn affect the inferred black hole
mass function. Additionally, accurate mass measurements of black holes
at the upper-end of the $M_\mathrm{BH} - \sigma_\star$ relationship
are necessary in order to address questions about whether the
$M_\mathrm{BH} - \sigma_\star$ or $M_\mathrm{BH} - L$ relationship is
more fundamental \citep{Lauer_2007}.

Recent work has suggested that some black hole masses at the high-mass
end of the $M_\mathrm{BH} - \sigma_\star$ and $M_\mathrm{BH} - L$
relationships may have been underestimated in previous stellar
dynamical measurements. In the case of M87, which contains one of the
largest measured black holes, \cite{Gebhardt_2009} found that
including the galaxy's dark matter halo in the stellar-dynamical
modeling increased the inferred black hole mass by a factor of $\sim
2$ to $6.4 \times 10^9\ M_\odot$. \cite{Shen_2010} re-examined the
stellar kinematics of NGC 4649, another galaxy harboring a high-mass
black hole, and found a factor of $\sim 2$ increase to $4.5 \times
10^9\ M_\odot$ from the original stellar-dynamical mass
measurement. Unlike M87, the addition of a dark matter halo had a
minor effect on the black hole mass. Instead, the mass increase was
attributed to the different orbital sampling -- the orbital coverage
from the previous model did not adequately sample the phase space
occupied by tangential orbits. Also, \cite{RvandenBosch_2010}
investigated the effects of using triaxial stellar-dynamical models
for a couple of objects that were originally modeled as axisymmetric
systems. Using a triaxial geometry in place of an axisymmetric shape
had a significant effect on one of the objects (NGC 3379), and the
black hole mass increased by a factor of $\sim 2$ to $4 \times 10^8\
M_\odot$. While the black hole in NGC 3379 does not fall at the
high-mass end of the $M_\mathrm{BH} - \sigma_\star$ and $M_\mathrm{BH}
- L$ relationships, \cite{RvandenBosch_2010} note that triaxial
systems may be prevalent at the high-end of the relationships.

These issues provide a renewed motivation to pursue gas-dynamical
measurements in massive, early-type galaxies.  Since gas-dynamical
measurements of black hole masses rely on gas in circular orbits at
small radii, within the black hole's dynamical sphere of influence,
they are insensitive to the large-scale effects of a dark matter halo
or stellar orbital anisotropy.  Thus, they can serve as an important
cross-check to the much more complex stellar-dynamical models.

The spectra of M84 illustrated by \cite{Bower_1998} show that the
black hole's dynamical sphere of influence is extremely well resolved
in the STIS data; this is demonstrated by the nearly Keplerian falloff
in rotation velocity with distance from the nucleus.  This makes M84 a
particularly valuable object for constraining the upper end of the
$M_\mathrm{BH} - \sigma_\star$ relation.  The discrepancy between the
masses derived by \cite{Bower_1998} and by
\cite{Maciejewski_Binney_2001} has remained troubling, and there are a
variety of possible explanations.  The modeling of \citet{Bower_1998}
was based on the decomposition of the emission-line profiles into
rotating and non-rotating components, but the models calculated by
\cite{Maciejewski_Binney_2001} clearly illustrated how a rotating disk
observed through a wide slit could give rise to the complex line
profiles seen in M84.  Also, the models calculated by
\cite{Bower_1998} did not include all of the detailed effects relevant
to propagation of emission-line profiles through the STIS optics that
were later explored by \cite{Maciejewski_Binney_2001},
\citet{Barth_2001}, \cite{Marconi_2003}, and others; these effects
could plausibly have a substantial impact on the derived black hole
mass.  On the other hand, the mass derived by
\cite{Maciejewski_Binney_2001} was based on the visual identification
of a caustic feature in the velocity field and not from quantitative
fits of models to the data, hence the uncertainty in their estimate of
$M_\mathrm{BH}$ is difficult to gauge.

Furthermore, neither \citet{Bower_1998} nor
\cite{Maciejewski_Binney_2001} included any possible effects due to an
intrinsic velocity dispersion in the disk or asymmetric drift in their
modeling.  Many nuclear gas disks in early-type galaxies exhibit an
internal velocity dispersion that may be dynamically important
\citep[e.g.,][]{vanderMarel_vandenBosch_1998, VerdoesKleijn_2002,
Neumayer_2007}. The physical origin of the intrinsic velocity
dispersion is not understood, and early gas-dynamical models did not
include the possible effect of the intrinsic velocity dispersion on
the black hole mass. In contrast, more recent gas-dynamical models
account for the possibility that the intrinsic velocity dispersion
provides dynamical support to the disk \citep{Barth_2001,
Coccato_2006, Neumayer_2007}.  The extremely broad emission lines near
the center of the M84 disk suggest that there could be a dynamically
significant internal velocity dispersion, which could have a
substantial impact on the measured black hole mass.

In this paper, we aim to resolve the uncertainty in the M84 black hole
mass by revisiting the archival STIS observations and carrying out
more comprehensive dynamical modeling than has previously been
attempted for this galaxy. We describe the archival \emph{HST} STIS
observations in \S \ref{sec:obs} and the measurement of the emission
lines in \S \ref{sec:measurement}. We then present the observed
emission-line velocity, velocity dispersion, and flux in \S
\ref{sec:velfields}. In \S \ref{sec:model}, we discuss the specifics
of our gas-dynamical model. We provide details of the stellar mass
profile, the emission-line flux distribution used in the model, the
contributions from various sources to the line widths, and the
asymmetric drift correction. The final disk models (both with and
without an asymmetric drift correction) are presented in \S
\ref{sec:modelresults}, and we quantify the possible sources of
uncertainty in \S \ref{subsec:errors}. Finally, in \S
\ref{sec:discussion}, we compare our black hole mass measurement to
past results. Throughout the paper, we adopt a distance to M84 of 17.0
Mpc in order to be consistent with \cite{Bower_1998} and
\cite{Gultekin_2009}.

\section{Observations and Data Reduction}
\label{sec:obs}

M84 was observed under program GO-7124 (see \citealt{Bower_1998}) with
the STIS \texttt{52x0.2} aperture and no gaps between the three
adjacent slit positions. The slit was aligned at a position angle of
104\degr\ east of north, approximately perpendicular to the radio jet
and near the major axis of the gaseous disk. The CCD was read out in
an unbinned mode giving a wavelength scale of $0.554$ \AA\
pixel$^{-1}$ and a spatial scale of $0$\farcs$0507$ pixel$^{-1}$. The
G750M grating was used to provide coverage of $6295$--$6867$ \AA,
which includes the H$\alpha$, [\ion{N}{2}] $\lambda\lambda 6548,
6583$, and [\ion{S}{2}] $\lambda\lambda 6716, 6731$ emission
lines. The exposure times ranged from $2245$ to $2600$ s per slit
position, with two individual exposures at each position.

The data were reduced using the standard Space Telescope Science
Institute (STScI) pipeline. The pipeline includes dark and bias
subtraction, flat-field corrections, and the combining of subexposures
to reject cosmic rays. The data were flux and wavelength calibrated,
and rectified for geometric distortions. Before the geometric
rectification, we performed an additional cleaning step to remove any
hot pixels or cosmic rays that remained in the flux-calibrated images.


\begin{figure*}
\begin{center}
\epsscale{0.6}
\plotone{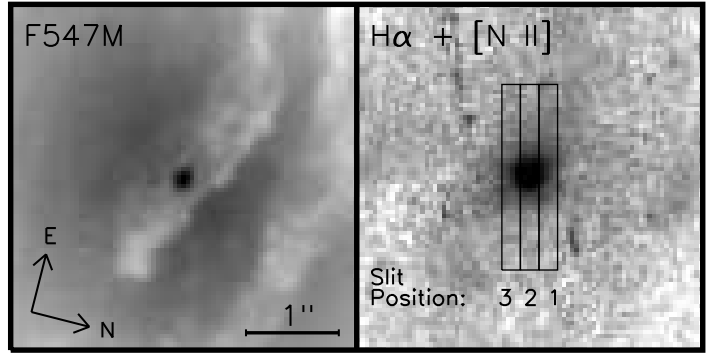}
\caption{\emph{HST} WFPC2/PC F547M (left) and continuum-subtracted
H$\alpha$+[\ion{N}{2}] (right) images of the M84 nucleus. STIS
observations were made at the three adjacent slit positions shown by
the rectangular regions overplotted on the continuum-subtracted
image. The length of the rectangular regions corresponds to the region
over which we could measure the emission lines. Both images have been
rotated such that the STIS instrumental $y$-axis points upward. Each
box is 3\farcs7 on a side. \label{fig:wfpc2image}}
\end{center}
\end{figure*}

We also obtained Wide Field Planetary Camera 2 (WFPC2) F547M, F658N,
and F814W images from the \emph{HST} archive, originally observed
under program GO-6094. In each image, the nucleus was centered on the
PC detector. The images for each of the filters were taken as a
sequence of two individual exposures, and we used the IRAF {\tt
combine} task to average the exposures together and reject cosmic
rays. However, a number of cosmic rays remained even after using the
{\tt combine} task, and so we applied an extra cosmic ray cleaning
step to the combined image using the LA-COSMIC task
\citep{vanDokkum_2001}. The total exposure times were 1200 s, 2600 s,
and 520 s for the F547M, F658N, and F814W images, respectively.

We created a continuum-subtracted H$\alpha$+[\ion{N}{2}] image by
subtracting a scaled combination of the F547M and F814W images from
the F658N image. We experimented with different scaling factors, and
searched for a set of factors that would produce a background region
in the continuum-subtracted image with a mean flux as close to zero as
possible. In Figure \ref{fig:wfpc2image}, we show the
continuum-subtracted image with the location of the STIS slits
overlaid, as well as the F547M WFPC2 image.  As previously described
by \citet{Bower_1997}, the emission-line image reveals a compact
central source surrounded by an extended disk-like structure that
traces the nuclear dust disk.

\section{Measurement of Emission Lines}
\label{sec:measurement}

We extracted spectra from individual rows of the 2D STIS image out to
about 0\farcs85 from the slit center. The rows were extracted as far
out as the emission lines were detectable. Far from the slit center,
where the emission lines were weak, we binned together multiple rows
to improve the S/N. Before fitting the emission lines, we removed the
continuum from the spectrum. For each row, we fit a line to the
continuum regions between rest-wavelengths $6500$--$6650$ \AA\ and
$6700$--$6750$ \AA, and then subtracted the continuum fit from the
spectrum. Generally, a straight line is not a good description of the
continuum near H$\alpha$, but with the low S/N in the continuum at
most positions and the small wavelength range, we could not perform a
more accurate subtraction.


\begin{figure*}
\begin{center}
\plotone{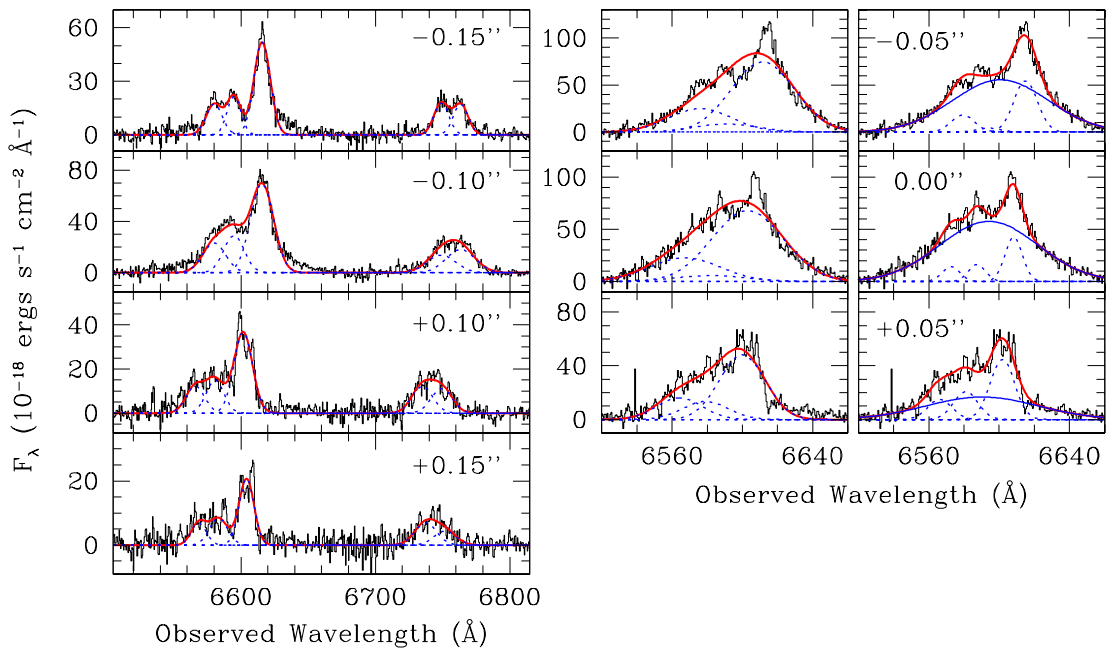}
\caption{Spectra extracted from the central slit located within
0\farcs15 from the M84 nucleus. For the spectra located at
$\pm$0\farcs1 and $\pm$0\farcs15 from the nucleus (\emph{left}), each
emission line was fit with a single Gaussian according to the basic
model described in \S \ref{sec:measurement}. For the spectra extracted
from the three innermost rows (\emph{right}), we show the fit results
from the basic model with an additional constraint on the H$\alpha$
and [\ion{N}{2}] line widths, and the same model but with a single
Gaussian broad component (shown by the blue solid line). In all
panels, the single Gaussian narrow components are shown by blue dashed
lines and the sum of all the Gaussian components are given by the red
solid line. \label{fig:spec_examples}}
\end{center}
\end{figure*}

We applied a Levenberg-Marquardt least-squares minimization technique
using the {\tt MPFIT} library in IDL \citep{Markwardt_2009} in order
to fit each spectrum with a set of Gaussians, following the same
technique described by \cite{Walsh_2008}. We used the propagated error
spectra to weight the data points during the fit. For all but the
three innermost rows of the central slit, we simultaneously fit five
Gaussians to the H$\alpha$, [\ion{N}{2}] $\lambda\lambda 6548, 6583$,
and [\ion{S}{2}] $\lambda\lambda 6716, 6731$ emission lines. All the
lines were required to have a common velocity, and the fluxes of the
[\ion{N}{2}] lines were held at a $3:1$ ratio. The widths of the
[\ion{N}{2}] lines were required to be the same. Also, the
[\ion{S}{2}] lines were constrained to have equal velocity widths.

Spectra extracted from the central slit position at 0\farcs15 and
0\farcs10 from the nucleus clearly exhibit double-peaked line
profiles. Our basic model described previously provides a good fit to
the spectra at 0\farcs15 and 0\farcs10, but we also tried fitting two
Gaussians to each emission line (representing two kinematically
distinct gas components). While the two-component model fits the
spectra very well at these locations, the $\chi^2$ per degree of
freedom ($\chi^2_\nu$) decreased by less than 12\% compared to the
basic model.

The basic model, however, did not return adequate fits for the spectra
extracted from the innermost three rows of the central slit position,
which have a very complex, asymmetric shape with severely blended
H$\alpha$ and [\ion{N}{2}] lines that may be double-peaked. We then
applied an additional constraint to the basic model that required the
widths of the H$\alpha$ and [\ion{N}{2}] lines to be equal. Even with
this model, the fit was poor and it was impossible to obtain accurate
and unique mean velocities, velocity dispersions, or fluxes. The
velocity dispersion values were very large, ranging from $\sim 600$ --
900 km s$^{-1}$, and the best-fit Gaussian appeared to be
systematically shifted blueward of the peak [\ion{N}{2}] $\lambda
6583$ flux in all three rows. We also tried fitting a number of more
complicated models, such as adding a single Gaussian broad component
to the H$\alpha$ + [\ion{N}{2}] complex or fitting a two-component
model. These more complex, multi-component fits were not well
constrained by the data, often resulting in Gaussian components with
zero flux. Moreover, including a broad component in the fit usually
resulted in unreasonably small fluxes and line widths for the
H$\alpha$ and [\ion{N}{2}] narrow components, although these models
did seem to provide a better estimate of the mean velocity than the
previous fitting attempts.

Past work by \cite{Bower_1998} has interpreted these complex line
profiles as arising from two separate gas components. They fit two
Gaussian components at number of locations within 0\farcs3 from the
nucleus, and only used one of the two components in order to derive a
black hole mass. However, as described above, modeling by
\cite{Maciejewski_Binney_2001} and \cite{Barth_2001b} has shown that
the line profiles can be described by a rotating disk model rather
than requiring two dynamically distinct components. Our detailed
modeling further supports this interpretation and will be discussed in
\S \ref{sec:modelresults}. Thus, for the spectra located at 0\farcs15
and 0\farcs10, we use the fit results from the basic model, with one
Gaussian component for each narrow emission line.

In the three innermost CCD rows of the central slit position, the
complexity and blending of the line profiles prevents an accurate
decomposition into Gaussian components, and it is not clear whether a
broad H$\alpha$ component is even present.  The broad wings on the
H$\alpha$+[\ion{N}{2}] complex could plausibly result from either a
compact broad-line region, or from the rapid rotation of the
innermost, unresolved portion of the emission-line disk, or both.
Thus, we consider all measurements from these central three rows to be
unreliable, and we do not use them to constrain our dynamical models.
Since the line profiles in these central three rows are dominated by
PSF blurring and rotational broadening, they add relatively little
useful information to constrain the disk models, and we will show in
\S \ref{sec:modelresults} that the modeling results are insensitive to
the values of the observed velocity from these central three STIS
rows.  However, for reference in the remaining figures presented in
the paper, we continue to plot the mean velocity (measured from a fit
using the basic model along with a single Gaussian broad component and
a constraint requiring the widths of the narrow H$\alpha$ and
[\ion{N}{2}] lines to be equal). We do not show the velocity
dispersion or flux measurements from the central three rows in these
figures though because we were unable to obtain reasonable
measurements from any of our fitting attempts.

In Figure \ref{fig:spec_examples}, we show a few examples of the basic
model fit to the spectra extracted from the central slit position at
$\pm$0\farcs1 and $\pm$0\farcs15 from the nucleus. We also display the
fits to the spectra from the innermost three rows using the basic
model plus an additional line width constraint, as well as the basic
model along with a single Gaussian broad component and a line width
constraint on the H$\alpha$ and [\ion{N}{2}] narrow components.

In order to estimate the uncertainties on the free parameters, we used
a Monte Carlo technique. For each row, we generated new spectra by
adding random Gaussian noise to the original spectrum, where the
1$\sigma$ level of the perturbation was set by the residuals between
the best-fit model and the original spectrum. The new spectra were
then refit, and the 1$\sigma$ uncertainties were taken to be the
standard deviation of the distribution. This method resulted in
velocity and velocity dispersion errors that were typically 27\% and
40\% larger, respectively, than the formal uncertainties from the
best-fit model.

\section{The Observed Velocity Field}
\label{sec:velfields}

\begin{figure*}
\begin{center}
\epsscale{1.1}
\plotone{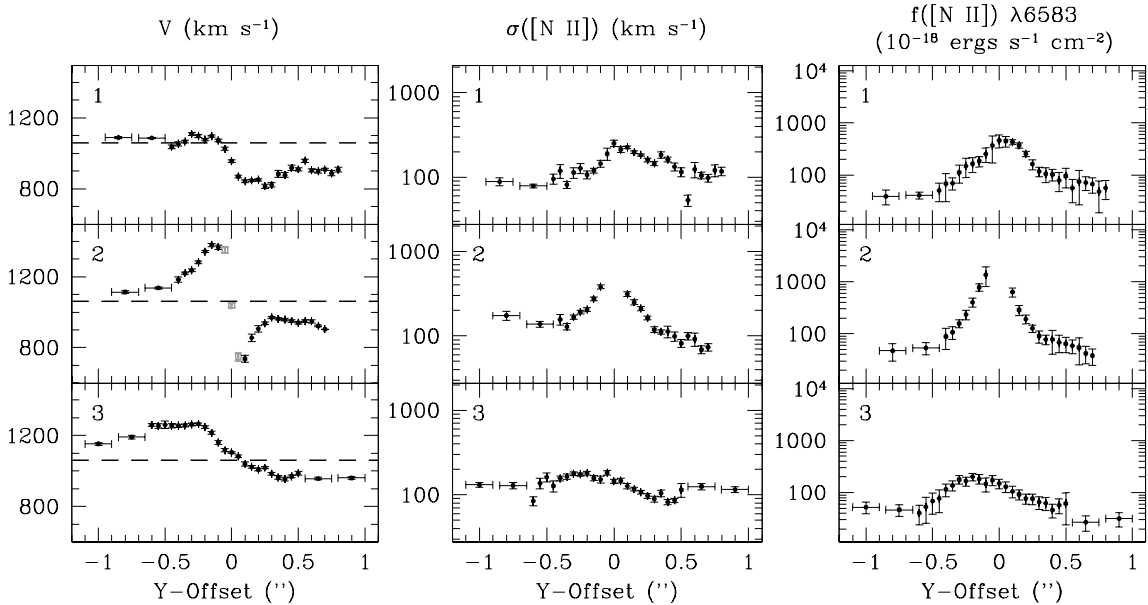}
\caption{The velocity, velocity dispersion, and flux measured from the
[\ion{N}{2}] $\lambda 6583$ emission line as a function of location
along each of the three slit positions in M84. The grey open squares
mark the velocity measurements from the three central rows where we
had difficulty fitting the spectra, while the unreliable velocity
dispersion and flux measurements from the three innermost rows are
left off the figure. The numbers at the top left of each plot
correspond to the slit locations depicted in Figure
\ref{fig:wfpc2image}. The positive and negative Y-Offsets correspond
to spectra extracted from the rows above and the rows below the
central row, respectively, when the slits are oriented such that the
STIS instrumental y-axis points upward. Thus, the bottom half of the
slit drawn in Figure \ref{fig:wfpc2image} corresponds to negative
Y-Offsets and the top half of the slit in Figure \ref{fig:wfpc2image}
corresponds to positive Y-Offsets. Spectral rows that were binned
together in order to improve the S/N are shown with error bars in the
Y-Offset direction. For reference, the dashed line shows the best-fit
systemic velocity determined from our gas-dynamical
modeling. \label{fig:velplots_dataonly}}
\end{center}
\end{figure*}

From the fits to the [\ion{N}{2}] $\lambda 6583$ emission line, we
obtained the velocity, velocity dispersion, and flux, and we plot
these values as a function of position along the slit in Figure
\ref{fig:velplots_dataonly}. We defined the galaxy center (Y-Offset =
0\arcsec) to be the row with the largest continuum flux, which also
coincided with the peak of the narrow [\ion{N}{2}] $\lambda 6583$
emission-line flux. The grey open squares mark the three central rows
where the complex line profiles precluded any definitive measurements
of the [\ion{N}{2}] mean velocity, velocity dispersion, or flux. These
grey points mark the velocity of the [\ion{N}{2}] emission peak as
determined by the fits that included a broad H$\alpha$ component. For
reference, we also show the best-fit systemic velocity determined from
our gas-dynamical modeling.


The multi-slit velocity curves show the gas participates in regular
rotation. There is a steep velocity gradient across the inner 0\farcs2
of slit position 2, where the radial velocity drops from 1380 km
s$^{-1}$ at Y-offset = $-$0\farcs1 to 740 km s$^{-1}$ at Y-offset =
0\farcs1. The radial velocities measured from slit position 3 show
that the gas located to the southwest of the nucleus is mostly
redshifted relative to the galaxy's systemic velocity, while the data
from slit position 1 show that the gas to the northeast of the nucleus
is mostly blueshifted. The [\ion{N}{2}] $\lambda 6583$ line widths
measured from the central slit position are large, with velocity
dispersions of $\sim 400$ km s$^{-1}$ and $\sim 300$ km s$^{-1}$ at
Y-offset = $-$0\farcs1 and 0\farcs1, respectively. The line widths
continue to rise toward the nucleus. However, due to the complexity of
the spectra, which are severely blended together due to the large
rotational broadening and the PSF blurring, we were unable to obtain
reliable measurements of the velocity, velocity dispersion, or flux
from the innermost three STIS rows.

\section{Modeling the Velocity Field}
\label{sec:model}

The velocity field was modeled assuming that the gas is in circular
rotation in a thin disk-like structure.  We follow a method similar to
that outlined by \cite{Barth_2001}, and we refer the reader there for
a more detailed discussion of the model.  At each radius $r$ in the
disk, we determine the circular velocity relative to the systemic
velocity ($v_\mathrm{sys}$), based on the enclosed mass $M(r)$, which
depends on the black hole mass ($M_\mathrm{BH}$), the stellar mass
profile, and the stellar mass-to-light ratio ($\Upsilon$).  The
stellar contribution to the circular velocities will be discussed in
\S \ref{subsec:stars}.

We then project the disk velocity field onto the plane of the sky for
a given value of the disk inclination $i$ in order to determine the
line-of-sight projection of the rotation velocity at each position in
the inclined disk.  We generated the model velocity field on a highly
subsampled pixel grid; each model pixel was 10$\times$ oversampled
relative to the STIS pixel size. Previous studies have used smaller
subsampling factors of $s=2$ and $s=4$ \citep{Barth_2001,
Coccato_2006, Wold_2006}.  However, we found that a large subsampling
factor was necessary in order to sufficiently capture the considerable
changes in velocity that occur over small spatial scales near the M84
nucleus, and to more accurately model the steep emission-line flux
profile. We experimented with a range of subsampling factors ($s$),
which will be discussed further in \S \ref{sec:modelresults}, before
settling on a subsampling factor of $s=10$.

We calculate the intrinsic line-of-sight velocity profiles on a
velocity grid with a bin size that matches the STIS pixel scale of
25.2 km s$^{-1}$. The intrinsic line-of-sight velocity profiles are
assumed to be Gaussian before passing through the telescope
optics. The Gaussian profiles are centered on the projected
line-of-sight velocity at each point on the model grid. Furthermore,
the line-of-sight velocity profiles are weighted by the emission-line
flux distribution (to be discussed in \S
\ref{subsec:emissionline_sb}). The width of the intrinsic
line-of-sight velocity profiles (to be discussed in \S
\ref{subsec:linewidths}) includes contributions from the thermal
velocity dispersion of the gas, the instrumental line spread function
(LSF) for STIS, and an intrinsic turbulent velocity dispersion.

The model velocity field is then synthetically ``observed'' in a
manner that matches the STIS observations. This synthetic observation
includes accounting for the blurring by the telescope PSF. We used
Tiny Tim \citep{Krist_Hook_2004} to create a 10$\times$ oversampled
0\farcs3$\times$0\farcs3 portion of the full STIS PSF for a
monochromatic filter passband at 6600 \AA. Although a
0\farcs3-diameter PSF model excludes part of the STIS PSF wings, there
is a negligible effect on the inferred black hole mass, as will be
demonstrated in \S \ref{sec:modelresults}. Each velocity slice of the
line profile grid is then convolved with the oversampled
0\farcs3$\times$0\farcs3 Tiny Tim PSF model.

After accounting for the telescope PSF, we propagate the model
velocity field through the STIS slit. The STIS slit is allowed to lie
at some angle $\theta$ from the projected major axis of the gas disk,
and the slit can be displaced some distance along the slit width
($x_\mathrm{offset}$) and along the slit length ($y_\mathrm{offset}$)
from the black hole. We included the shifts in velocity that arise as
a result of the finite slit width. The velocity shifts occur because a
photon's recorded wavelength depends on the location along the slit
width at which the photon enters \citep[e.g.,][]{Barth_2001,
Maciejewski_Binney_2001}.

After the velocity shifts have been made, we rebin the resulting
emission-line profiles to the STIS pixel size. Thus, we are left with
a model 2D spectral image similar to the STIS data. We convolve the
model 2D spectral image with the CCD charge diffusion kernel. The CCD
charge diffusion kernel is given by Tiny Tim for model PSFs that are
subsampled, and it represents the charge that is spread between the
immediate neighboring (non-subsampled) pixels.

Finally, spectra are extracted on a row-by-row basis from the model
STIS 2D image. We fit a single Gaussian to the emission line,
analogous to the measurements of the emission lines from the STIS
data. We are thus able to measure the model velocity, velocity
dispersion, and flux as a function of position along the slit. We then
determine the best-fit model parameters ($M_\mathrm{BH}$, $\Upsilon$,
$i$, $v_\mathrm{sys}$, $\theta$, $x_\mathrm{offset}$,
$y_\mathrm{offset}$) that produce a model velocity field that most
closely matches the observed velocity field over a region
($r_\mathrm{fit}$) that extends from $-$0\farcs5 to 0\farcs5 for each
slit position. However, we do not include the uncertain velocity
measurements from the innermost three STIS rows in the fit.  We
measure the black hole mass and the $\chi^2$ from the model fits to
the observed velocity only, after separately optimizing the model fits
to the observed line widths and fluxes as described below. The
observed [\ion{N}{2}] line widths and fluxes are not directly
dependent on the black hole mass, so the black hole mass is determined
from the fit to just the radial velocity curves.

\subsection{Stellar Mass Profile}
\label{subsec:stars}

In order to account for the stellar contribution to the gravitational
potential, we modeled the intrinsic density distribution of the
central regions of M84 as the sum of spherically symmetric components
with Gaussian profiles, as done by \cite{Sarzi_2001},
\cite{Barth_2001}, \cite{Sarzi_2002}, and \cite{Coccato_2006}. Within
such a multi-Gaussian framework \citep{Monnet_1992}, projecting the
model density profile on the sky to match the observed surface
brightness is fairly simple given the properties of the Gaussian
function. This is particularly true when the instrumental PSF is also
expressed in terms of Gaussian functions. Once the (non-negative)
amplitudes of the Gaussian density components have been determined,
the contribution of each Gaussian to the stellar potential can be
conveniently evaluated in terms of error functions.

For M84, we deprojected the $V$-band surface brightness profile of
\cite{Kormendy_2009}, who combined large scale ground-based images
with high-resolution \emph{HST} NICMOS $H$-band data for the central,
dusty region. Since the NICMOS PSF is fairly complex, we took
particular care to describe it in terms of Gaussian components by
matching to a NIC2 F205W PSF generated with Tiny Tim. A careful
representation of the PSF is important in the presence of an active
nucleus, as in the case of M84, because it affects the isolation and
removal of an AGN component from the stellar mass budget.

The luminosity profile of M84 contains a compact central component.
\citet{Bower_2000} show that this central component is due to the AGN
itself and not a nuclear star cluster, so the light from this
component should not contribute to the stellar mass profile. In order
to determine the impact of this component on the mass models, we
computed two different deprojected stellar mass profiles, one in which
this central component was assumed to be AGN light, and another in
which it was assumed to be starlight with the same stellar
mass-to-light ratio as the rest of the galaxy.

\begin{figure}
\begin{center}
\epsscale{1.1}
\plotone{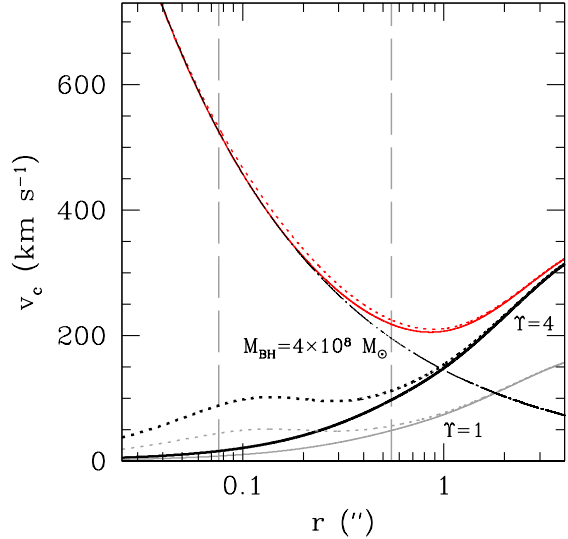}
\caption{Stellar contribution to the circular velocity ($v_c$)
obtained from the multi-Gaussian deprojection of the M84 $V$-band
surface brightness profile. The thick black and thin grey radially
increasing lines show $v_c$ values obtained assuming stellar
mass-to-light ratios of $\Upsilon=4$ and $1$, respectively.  Two
different stellar $v_c$ values are shown; the solid curves illustrate
the case where the central compact source in the light profile is
assumed to be the AGN (in which case it makes no contribution to the
stellar mass profile), while the dotted curves illustrate the case in
which it is assumed to be a compact stellar nucleus.  The black
dot-dashed line shows the Keplerian velocities produced by a $4 \times
10^8\ M_{\odot}$ black hole, which, when added to the stellar
potential obtained with $\Upsilon=4$, would lead to the red solid and
dotted lines that trace a nearly parabolic trend. Over the radial
range we fit the kinematic data (shown by the vertical grey dashed
lines, see \S \ref{sec:modelresults}), $v_c$ is almost entirely
determined by the black hole mass.  Whether the light from the central
compact component is excluded or included in the mass budget does not
significantly alter the estimate of $M_{\rm
BH}$. \label{fig:stellarvc}}
\end{center}
\end{figure}

In Figure \ref{fig:stellarvc}, we present the final product of the
multi-Gaussian deprojection of the surface-brightness profile of M84,
that is, the stellar contribution to the circular velocity in the
galaxy.  The circular velocity due to a $4\times10^8$ $M_\odot$ black
hole is also illustrated, since this is a plausible lower limit to the
black hole mass based on our results and those of
\citet{Maciejewski_Binney_2001}.  The figure illustrates that the
stellar contribution to the circular velocities is extremely small
over the radial range probed by our data ($\sim0\farcs5$), even for
the lowest reasonable value of $M_{\rm BH}$.  The stellar contribution
is negligibly small regardless of whether the central compact
component in the light profile is assumed to be a star cluster or an
AGN.

The results shown in Figure \ref{fig:stellarvc} indicate that the
total mass within the inner 0\farcs5 is likely to be dominated by the
black hole.  In other words, the black hole's dynamical sphere of
influence is extremely well resolved by the STIS observations; this is
consistent with the nearly Keplerian shape of the velocity profile
along the central slit position.  As a result, our model fits should
be fairly insensitive to the stellar mass-to-light ratio $\Upsilon$.
In fact, as we describe below, our dynamical models do not provide any
useful constraints on $\Upsilon$, and we simply assume a fixed value
of $\Upsilon$ for our final model fits.  In our dynamical models, we
used the circular velocities derived under the assumption that the
central compact component represents AGN light.


\subsection{Emission-Line Flux}
\label{subsec:emissionline_sb}

As discussed above in \S \ref{sec:model}, the model line-of-sight
velocity profiles are weighted by the emission-line flux at each point
in the disk. Past gas-dynamical models have used either an analytic
form to describe the intrinsic emission-line flux
\citep[e.g.,][]{VerdoesKleijn_2002, Coccato_2006, Marconi_2006,
Pastorini_2007, Hicks_Malkan_2008} or the continuum-subtracted
H$\alpha$ + [\ion{N}{2}] image directly, after deconvolution with the
telescope PSF \citep[e.g.,][]{Barth_2001, Shapiro_2006, Wold_2006,
DallaBonta_2009}. \cite{Barth_2001} found that folding the deconvolved
continuum-subtracted image into the model calculations can account for
small-scale irregularities in the velocity field. However, our
attempts to use a deconvolved image of M84 were unsuccessful, and the
dynamical model produced poor fits to both the observed flux and
velocity. It appeared that the deconvolved continuum-subtracted image
could not adequately characterize the steepness of the nuclear
emission-line light on subpixel scales.

We therefore experimented with analytic forms of various complexity
for the intrinsic flux distribution. We searched for the simplest
function that would sufficiently reproduce the observed [\ion{N}{2}]
flux profile while simultaneously producing model velocities closest
to the observed velocities. We tested parametrizations that were
composed of two to four components, and the components were either
Gaussian functions [parametrized as $F(r) = F_0 \exp(-r^2 / 2r_s^2)$],
exponential functions [parametrized as $F(r) = F_0 \exp(-r / r_s)$],
or constants.  We experimented with models representing intrinsically
circularly symmetric disks (producing concentric elliptical isophotes
with constant position angle and axis ratio), as well as more
complicated models where the isophotes of the individual components
were allowed to have different centers, positions angles, and axis
ratios.  The parameters for each possible surface brightness model
were determined by computing disk models including the parametrized
surface brightness model, calculating the resulting model line
profiles as described above in \S \ref{sec:model}, measuring the
emission-line fluxes in the resulting model, and then optimizing the
fit of the modeled fluxes to the observed [\ion{N}{2}] flux
distribution by minimizing $\chi^2$.  Ultimately, we found that the
best parametrization was the sum of four components: three concentric
Gaussians describing the compact nucleus, plus a more extended
exponential component with an offset center relative to the positions
of the Gaussian components.  The amplitudes ($F_0$), scale radii
($r_s$) along the major axis, position angles, axis ratios ($b/a$),
and centroid positions for each of these elliptical components in the
best-fitting model are listed in Table \ref{tab:sbparms}.


\begin{deluxetable}{lcccccc}
\tabletypesize{\scriptsize} 
\tablewidth{0pt} 
\tablecaption{Emission-Line Flux Parameters \label{tab:sbparms}}
\tablehead{
\colhead{Component} & 
\colhead{$F_0$} & 
\colhead{$r_s$} &
\colhead{$x_\mathrm{cen}$} &
\colhead{$y_\mathrm{cen}$} &
\colhead{PA} &
\colhead{$b/a$} \\
\colhead{} &
\colhead{} &
\colhead{(pc)} &
\colhead{(\arcsec)} &
\colhead{(\arcsec)} &
\colhead{($^{\circ}$)} &
\colhead{}
}

\startdata

Gaussian     &  234.2  &   0.5  &  0.00  &   0.00  &  344  &  0.9  \\
Gaussian     &  126.6  &   2.8  &  0.00  &   0.00  &   33  &  0.4  \\
Gaussian     &    6.0  &  12.0  &  0.00  &   0.00  &   31  &  0.4  \\
Exponential  &    1.0  &  50.0  &  0.13  &   0.02  &   69  &  0.7  \\

\enddata

\tablecomments{The amplitude, $F$, is in arbitrary flux units. The
center of the ellipse is described by $x_\mathrm{cen}$ and
$y_\mathrm{cen}$, relative to the location of the black hole. The
position angle, PA, is in units of degrees clockwise, with PA $= 0$
pointing along the length of the STIS slit.}

\end{deluxetable}

\subsection{Line Widths}
\label{subsec:linewidths}

A number of factors contribute to the width of the line-of-sight
velocity profiles, such as rotational broadening, instrumental
broadening, thermal broadening, and possibly an intrinsic velocity
dispersion in the gaseous disk. Rotational broadening occurs because
light from different parts of the disk falls within the same slit and
consequently is blended together. Instrumental broadening includes the
PSF blurring, the velocity shifts that occur as a result of the finite
slit width, and the effects of charge diffusion between neighboring
pixels. All of these instrumental effects, as well as the rotational
broadening, are explicitly included in our modeling when the intrinsic
line-of-sight velocity profiles are propagated through the telescope
and spectrograph. Before the line-of-sight velocity profiles are
propagated through the telescope optics, we assign the line profiles a
small width resulting from the intrinsic instrumental line-spread
function estimated by \cite{Barth_2001} to be $\sigma_{\mathrm{LSF}} =
8$ km s$^{-1}$ and a velocity dispersion of $\sigma_{\mathrm{th}} =
10$ km s$^{-1}$, which is the expected thermal contribution to the
line width for gas with a temperature of $\approx 10^4$ K. Since the
values of $\sigma_{\mathrm{LSF}}$ and $\sigma_{\mathrm{th}}$ are very
small compared with the observed line widths, they have a negligible
effect on the model calculation.

Even though the models accounted for the rotational, instrumental, and
thermal broadening, our preliminary models predicted line widths that
were much smaller than the observed velocity dispersions in M84. This
behavior has often been seen in past studies of other galaxies
\citep[e.g.,][]{vanderMarel_vandenBosch_1998, VerdoesKleijn_2000,
VerdoesKleijn_2002, Shapiro_2006, DallaBonta_2009}. However, previous
work has also found that rotational and instrumental broadening alone
can explain the observed line widths in some objects (e.g.,
\citealt{Macchetto_1997, Capetti_2005, Atkinson_2005,
deFrancesco_2006}, 2008). In order to produce model line widths that
match the observations, we include a projected intrinsic velocity
dispersion of the form

\begin{equation}
\label{eq:sigmar}
\sigma_\mathrm{p} = \sigma_0 + \sigma_1 \exp{ \bigg( \frac{-r}{r_0}
\bigg) } \;.
\end{equation}

\noindent{Thus, the width of the line-of-sight velocity profiles is
given by $\sigma_{\mathrm{th}}$, $\sigma_{\mathrm{LSF}}$, and
$\sigma_\mathrm{p}$ added in quadrature, combined with the width
resulting from the propagation of the line profiles through the
telescope and spectrograph. We determined that $\sigma_0 = 77$ km
s$^{-1}$, $\sigma_1 = 214$ km s$^{-1}$, and $r_0 = 26$ pc by fitting
the line widths predicted from a preliminary model to the observed
line widths. The unreliable velocity dispersion measurements from the
three central STIS rows were excluded from the fit.}

The physical nature of the intrinsic velocity dispersion is not
understood. \cite{vanderMarel_vandenBosch_1998} proposed that the
intrinsic velocity dispersion is the result of local microturbulence,
but that the bulk motion of the gas remains in circular motion. Others
suggest the local random motions may contribute pressure support to
the disk, which if ignored will lead to an underestimate of the black
hole mass \citep[e.g.,][]{Barth_2001,
Neumayer_2007}. \cite{VerdoesKleijn_2006} find that radio galaxies in
particular (including M84) often show evidence for nuclear velocity
dispersions in excess of those expected from pure gravitational
motions. They attribute the excess velocity dispersion to
non-gravitational motions in the gas.

While there remains no current clear consensus as to the physical
mechanism that causes the discrepancy between the model line widths
and the observed velocity dispersions, we examine two scenarios that
should cover the possible range of masses for the black hole in
M84. First, we consider the case in which the intrinsic velocity
dispersion does not affect the circular velocity. Secondly, we analyze
the case in which the intrinsic velocity dispersion contributes
dynamical support to the disk, and we apply an asymmetric drift
correction.

\subsection{Asymmetric Drift Correction}
\label{subsec:asymmdrift}

If the intrinsic velocity dispersion provides pressure support that
balances gravity in the gas disk, then the observed mean rotation
speed ($v_\phi$) will be smaller than the local circular velocity
($v_c$) for a given black hole mass. Models that do not account for
this effect will lead to black hole mass measurements that
underestimate the true mass. Although it is unclear whether the
intrinsic velocity dispersion actually does provide dynamical support
to the disk, we include an asymmetric drift correction in our model to
estimate an upper limit to the M84 black hole mass, following the
methods previously described by \cite{Barth_2001}.

Briefly, if we assume that the gas motions are isotropic in the $r$
and $z$ (cylindrical) coordinates, then asymmetric drift correction is
given by

\begin{equation}
\label{eq:asymmdriftcorr}
v_c^2\ -\ v_\phi^2\ =\ \sigma_r^2 \bigg[ -\frac{d \ln \nu}{d \ln r}\ -\ 
\frac{d \ln \sigma_r^2}{d \ln r}\ -\ 
\bigg( 1\ -\ \frac{\sigma_\phi^2}{\sigma_r^2} \bigg) \bigg] \;.
\end{equation}

\noindent Here, $\nu$ is the number density of tracer particles or
clouds in the gaseous disk, and we assume that intrinsic radial
velocity dispersion $\sigma_r$ can be described by an exponential +
constant of the same form as Equation \ref{eq:sigmar}. For a given
parametrization of $\sigma_r$, we determine $\sigma_\phi$ and the
projected velocity dispersion following the method described by
\cite{Barth_2001}. This projected velocity dispersion is then added in
quadrature to $\sigma_\mathrm{th}$ and $\sigma_\mathrm{LSF}$
(discussed in \S \ref{subsec:linewidths}) in order to obtain the
widths of the line-of-sight velocity profiles before propagation
through the telescope and spectrograph. We found best-fit values of
$\sigma_0 = 37$ km s$^{-1}$, $\sigma_1 = 255$ km s$^{-1}$, and $r_0 =
53$ pc by computing preliminary disk models with an asymmetric drift
correction and fitting the model line widths to the observed
[\ion{N}{2}] line widths. A similar expression for the asymmetric
drift correction is given by \cite{Valenzuela_2007}, including an
additional term that accounts for the effect of gas pressure
gradients. We neglect this additional term as it provides only a small
contribution to the asymmetric drift correction.

\section{Modeling Results}
\label{sec:modelresults}

In order to determine the best-fit model parameters ($M_\mathrm{BH}$,
$\Upsilon$, $i$, $v_\mathrm{sys}$, $\theta$, $x_\mathrm{offset}$,
$y_\mathrm{offset}$), we minimized $\chi^2$ using the downhill simplex
algorithm by \cite{Press_1992}. Initially, we ran preliminary models
without an asymmetric drift correction allowing all seven model
parameters to vary independently. However, it became clear that the
stellar mass-to-light ratio could not be well constrained by the STIS
data. The black hole's sphere of influence is so well resolved that we
are unable to determine a precise value for $\Upsilon$. Our
preliminary models often returned unreasonably low values of $\Upsilon
= 0$. Therefore, we estimated $\Upsilon$ based on the $B-V$ color for
the galaxy using \cite{Bell_2003}. The $B-V$ color from the Third
Reference Catalogue of Bright Galaxies (RC3) \citep{RC3_catalog} is
0.98, which implies $\Upsilon \sim 4$ in $V$-band solar units. We then
fixed $\Upsilon$ to this value in our models.

\begin{figure*}
\begin{center}
\plotone{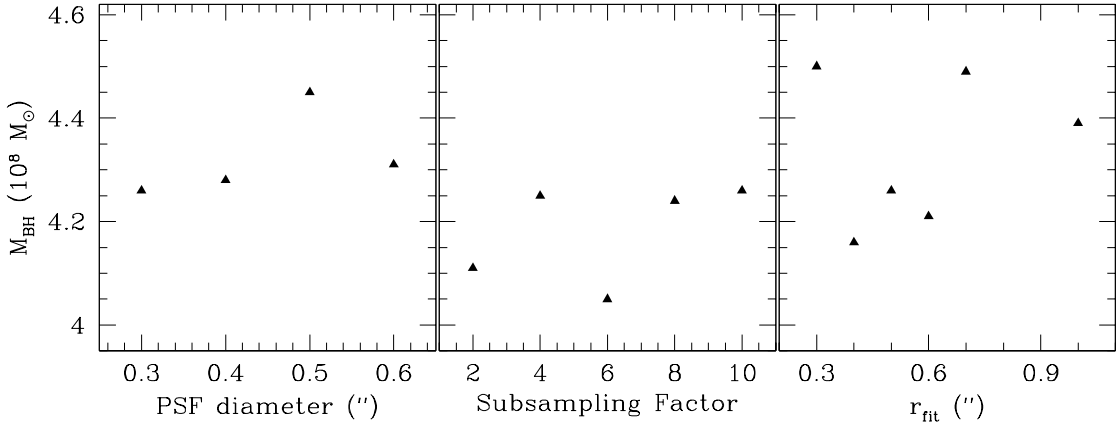}
\caption{Black hole mass as a function of PSF size (with $s=10$ and
$r_\mathrm{fit}= $ 0\farcs5), subsampling factor (with a
0\farcs3-diameter PSF and $r_\mathrm{fit}=$ 0\farcs5), and fitting
radius (with a 0\farcs3-diameter PSF and $s=10$), respectively. The
quantities $M_\mathrm{BH}$, $i$, $\theta$, $v_\mathrm{sys}$,
$x_\mathrm{offset}$, and $y_\mathrm{offset}$ were allowed to vary
during the fit, but $\Upsilon$ was fixed at 4 $V$-band solar
units. All models were calculated without an asymmetric drift
correction. \label{fig:modeltests}}
\end{center}
\end{figure*}

We also calculated preliminary models using different PSF sizes,
subsampling factors, and fitting regions. During all of these model
calculations, the parameters $M_\mathrm{BH}$, $i$, $\theta$,
$v_\mathrm{sys}$, $x_\mathrm{offset}$, and $y_\mathrm{offset}$ were
allowed to vary during the fit, but $\Upsilon$ was fixed at 4 in
$V$-band solar units. Also, the uncertain velocity measurements from
the three central STIS rows were excluded from the fit. In Figure
\ref{fig:modeltests}, we present the results of the models without an
asymmetric drift correction.


In order to test the significance of the PSF size on the black hole
mass measurement, we calculated disk models using Tiny Tim PSFs
ranging in size from 0\farcs3 to 0\farcs6 in diameter. All of these
model PSFs exclude part of the extended STIS PSF wings, with the
omitted flux totaling 18\%, 13\%, 11\%, and 9\% of the entire PSF flux
for a 0\farcs3, 0\farcs4, 0\farcs5, and 0\farcs6-diameter PSF,
respectively. We found that the PSF size has a small impact on the
black hole mass measurement, with $M_\mathrm{BH}$ varying between $4.3
\times 10^8\ M_\odot$ and $4.5 \times 10^8\ M_\odot$ for the models
without an asymmetric drift correction. Because the computation time
increases rapidly with PSF size, and the PSF size has a minimal impact
on the black hole mass, we used the Tiny Tim 0\farcs3-diameter PSF in
our final models.

When testing the effect of the subsampling factor on the black hole
mass, we calculated disk models using using subsampling factors
between $s=2$ and $s=10$ in increments of two. We found that the black
hole mass did not vary significantly with the subsampling factor. For
models without an asymmetric drift correction, $M_\mathrm{BH}$
fluctuated between $4.1 \times 10^8\ M_\odot$ and $4.3 \times 10^8\
M_\odot$. Although the black hole mass was almost unaffected by the
subsampling factor, we found that larger subsampling factors resulted
in much better fits to both the emission-line flux and
velocities. Small subsampling factors were unable to capture the large
changes in both the velocity and emission-line flux seen near the M84
nucleus. Thus, in our final models we used $s=10$.

Similarly, we calculated disk models with $r_\mathrm{fit}$ values
between 0\farcs3 and 1\arcsec. Like with the PSF and subsampling
factor, the size of the fitting region had a small impact on the black
hole mass, with $M_\mathrm{BH}$ between $4.2 \times 10^8\ M_\odot$ and
$4.5 \times 10^8\ M_\odot$ for the models without an asymmetric drift
correction. Using the largest possible fitting region is not
necessarily the best approach because there can be systematic
departures from the disk model at large radii, possibly due to disk
warping.  The region in the central slit position near
Y-Offset$=-0\farcs5$, in particular, is not well fit by any of the
disk models. We used $r_\mathrm{fit} = $ 0\farcs5 in our final models
because this region is large enough to contain an adequate number of
velocity measurements needed to constrain the model, but also small
enough so that the disk model still provides a good description of the
observations.

\begin{figure*}
\begin{center}
\plotone{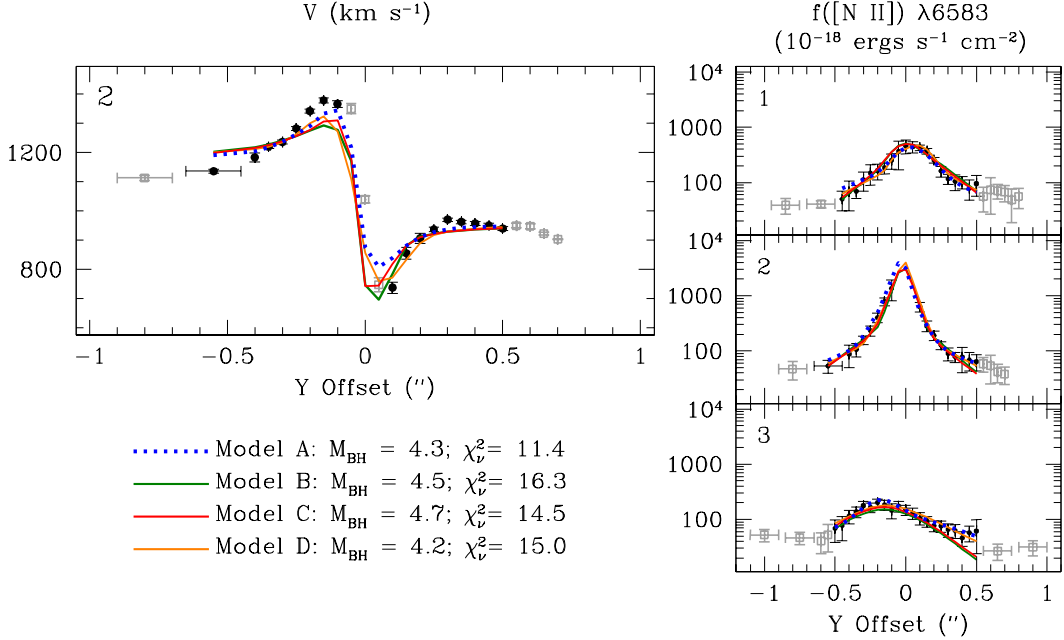}
\caption{Results of disk models with four different parametrizations
of the emission-line flux: 3 Gaussians $+$ 1 exponential (model A), 1
Gaussian $+$ 1 exponential (model B), 2 exponentials (model C), and 2
Gaussians $+$ 1 exponential (model D). We present the model velocity
for the central slit position (left) and the model emission-line flux
from all three slit positions (right), as well as the best-fit
$M_\mathrm{BH}$ in units of $10^8\ M_\odot$ and the corresponding
$\chi^2_\nu$. All models were calculated without an asymmetric drift
correction using a 0\farcs3-diameter PSF and $s=10$. The fitting
radius was set to $r_\mathrm{fit} = $ 0\farcs5, and thus the black
filled circles are the data measured from the STIS observations that
were included in the fit while the grey open squares are the data that
were excluded. The quantities $M_\mathrm{BH}$, $i$, $\theta$,
$v_\mathrm{sys}$, $x_\mathrm{offset}$, and $y_\mathrm{offset}$ were
allowed to vary during the fit. \label{fig:sbtest}}
\end{center}
\end{figure*}

We also experimented with using different analytic parametrizations of
the intrinsic emission-line flux distribution. In Figure
\ref{fig:sbtest}, we present the results of disk models without an
asymmetric drift correction, in which the emission-line flux was
modeled with three additional functions of varying complexity: 1
Gaussian $+$ 1 exponential (model B), 2 exponentials (model C), and 2
Gaussians $+$ 1 exponential (model D). The isophotes of the components
in models B, C, and D are elliptical and have different centers,
position angles, and axis ratios similar to the emission-line flux
distribution composed of 3 Gaussians $+$ 1 exponential adopted for the
final model (model A) discussed in \S \ref{subsec:emissionline_sb}.


Models A, B, C, and D reproduced the observed emission-line flux
within the errors, but the quality of the fit to the observed velocity
varied significantly ($\chi^2_\nu$ ranged from 11.4 -- 16.3 for the
models without an asymmetric drift correction). Although the intrinsic
emission-line flux distribution affected the quality of the velocity
fit, it only had a small influence on $M_\mathrm{BH}$, with the
best-fit mass between $4.2 \times 10^8\ M_\odot$ and $4.7 \times 10^8\
M_\odot$. The observed emission-line flux could also be fit with two
additional analytic functions: 3 exponentials and 3 Gaussians $+$ 1
exponential with a nuclear hole. When optimizing this last model to
the observed flux distribution, we found a best-fit hole radius of 0.8
pc, which is well within a single STIS pixel. The emission-line flux
models composed of 3 exponentials and 3 Gaussians $+$ 1 exponential
with a nuclear hole produced a black hole mass of $4.5 \times 10^8\
M_\odot$ and a $\chi^2_\nu$ of 13.7 and 12.3, respectively. These two
models are left off Figure \ref{fig:sbtest} for clarity, but similarly
show that the black hole mass does not change drastically. Our
findings are consistent with \cite{Marconi_2006}, who demonstrated
that the emission-line flux model has little effect on the final black
hole mass measurement for Centaurus A. Many other emission-line flux
models were tested, but the models returned unacceptable fits to the
observed flux, and thus they are not presented here.

The other disk parameters appeared to be well constrained by the
multi-slit STIS data. Throughout the numerous calculations of
preliminary models without an asymmetric drift correction, the
parameters $i$, $v_\mathrm{sys}$, $\theta$, $x_\mathrm{offset}$, and
$y_\mathrm{offset}$ varied at most by 8$^{\circ}$, 10 km s$^{-1}$,
6$^{\circ}$, 0\farcs03, and 0\farcs03, respectively. Similarly, we
found deviations of at most 10$^{\circ}$, 12 km s$^{-1}$, 8$^{\circ}$,
0\farcs03, and 0\farcs04 for $i$, $v_\mathrm{sys}$, $\theta$,
$x_\mathrm{offset}$, and $y_\mathrm{offset}$, respectively, for the
models with an asymmetric drift correction. The uncertainty in the
black hole mass due to these parameters is negligible, but will be
accounted for in \S \ref{subsec:errors} when the formal model fitting
uncertainty is determined.

\subsection{Models Without an Asymmetric Drift Correction}
\label{subsec:results_noasymmdrift}


\begin{figure*}
\begin{center}
\epsscale{1.1}
\plotone{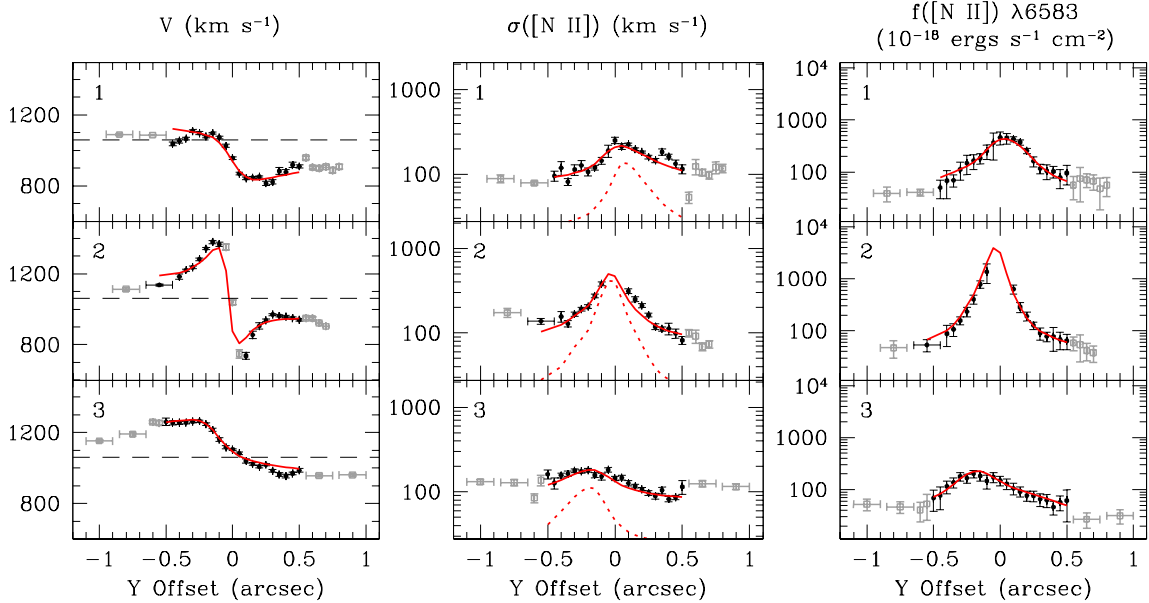}
\caption{Results of the final model without an asymmetric drift
correction using a 0\farcs3-diameter PSF, $s=10$, and
$r_\mathrm{fit}=$ 0\farcs5. The model velocity, velocity dispersion,
and emission-line flux are shown by the red solid line. The red dotted
line shows the model velocity dispersion if an intrinsic velocity
dispersion is not included in the model. We applied a single scaling
factor to the model emission-line flux such that the median values of
the model and the data would match. The black filled circles are the
data from the STIS observations that were included in the fit, while
the grey open squares are the data that were excluded from the
fit. For reference, the horizontal dashed line shows the best-fit
systemic velocity determined from the final
model. \label{fig:bestfitmodel_noasymmdrift}}
\end{center}
\end{figure*}

The final model without an asymmetric drift correction was calculated
using a 0\farcs3 $\times$ 0\farcs3 PSF, a subsampling factor of $s =
10$, and a fitting region of $r_\mathrm{fit} = $ 0\farcs5.  Due to the
complexity of the spectra extracted from the three central STIS rows
and the resulting uncertainty in the radial velocity measurements, we
excluded these three innermost measurements from the final fit. The
stellar mass-to-light ratio was fixed to a value of 4 in $V$-band
solar units. We also included the projected intrinsic velocity
dispersion described in \S \ref{subsec:linewidths}.

In Figure \ref{fig:bestfitmodel_noasymmdrift} we compare the final
model velocity, velocity dispersion, and emission-line flux to the
observed values as a function of location along the slit for each of
the three slit positions. As can be seen in this figure, the final
model is able to sufficiently reproduce the shape of the observed
velocity field and the emission-line flux. It is also apparent that an
intrinsic velocity dispersion is needed in order to bring the model
into agreement with the observations. In the left column of Table
\ref{tab:bestfitmodels}, we give the best-fit parameter values for the
final model without an asymmetric drift correction.

Even though the final model matches the general shape of the velocity
curves fairly well, we find that $\chi^2 = 594.7$. The 58 velocity
measurements are fit with 6 parameters ($M_\mathrm{BH}$, $i$,
$v_\mathrm{sys}$, $\theta$, $x_\mathrm{offset}$, and
$y_\mathrm{offset}$), resulting in $\chi^2_\nu = 11.4$. We acknowledge
that our thin-disk model is unable to reproduce all of the velocity
structure that is observed. As will be shown below in \S
\ref{subsec:errors}, removing several velocity measurements at
locations that clearly deviate from pure circular rotation leaves the
black hole mass unaffected while significantly improving
$\chi^2_\nu$. Thus, we believe our $M_\mathrm{BH}$ measurement remains
credible despite the moderate $\chi^2_\nu$ value. Furthermore, we
account for the imperfect disk model in \S \ref{subsec:errors} by
renormalizing $\chi^2_\nu$ before estimating the formal uncertainty in
$M_\mathrm{BH}$ due to the model fitting process. We use this approach
because it will provide the most conservative estimate of the
uncertainty in the black hole mass. We note that several gas-dynamical
models of other objects have similar $\chi^2_\nu$ values
\citep[e.g.,][]{Capetti_2005, Atkinson_2005, deFrancesco_2008,
DallaBonta_2009}.


\begin{deluxetable}{lcc}
\tabletypesize{\scriptsize} 
\tablewidth{0pt} 
\tablecaption{Final Disk Model Parameters \label{tab:bestfitmodels}}
\tablehead{
\colhead{} & 
\colhead{w/o Asymmetric} & 
\colhead{w/ Asymmetric} \\
\colhead{} & 
\colhead{Drift} & 
\colhead{Drift}
}

\startdata

$M_\mathrm{BH}$\ $(M_\odot)$      & (4.3$^{+0.8}_{-0.7}$)$\times$10$^8$ & (8.5$^{+0.9}_{-0.8}$)$\times$10$^8$ \\
$\Upsilon$\ ($V$-band solar)      & 4 & 4 \\
$i$\ ($^\circ$)                   & 67$^{+1}_{-7}$ & 72$^{+1}_{-9}$ \\
$v_\mathrm{sys}$\ (km s$^{-1}$)   & 1060$^{+6}_{-4}$ & 1060$^{+9}_{-3}$ \\
$\theta$ ($^\circ$)               & 27$\pm3$ & 28$^{+3}_{-5}$ \\
$x_\mathrm{offset}$\ (\arcsec)    & 0.02$^{+0.02}_{-0.01}$ & 0.01$^{+0.02}_{-0.01}$ \\
$y_\mathrm{offset}$\ (\arcsec)    & $-$0.04$^{+0.02}_{-0.01}$ & $-$0.05$^{+0.03}_{-0.01}$ \\

\enddata

\tablecomments{Uncertainties given for the black hole mass are 68\%
confidence limits. The stellar mass-to-light ratio, $\Upsilon$, was
frozen at 4 ($V$-band solar units) for both models. A relative angle
of $\theta = 27^\circ$ and $\theta = 28^\circ$ corresponds to a disk
major axis position angle of 77$^\circ$ and 76$^\circ$ east of north.}

\end{deluxetable}

\subsection{Models Including an Asymmetric Drift Correction}
\label{subsec:results_asymmdrift}


\begin{figure*}
\begin{center}
\epsscale{1.1}
\plotone{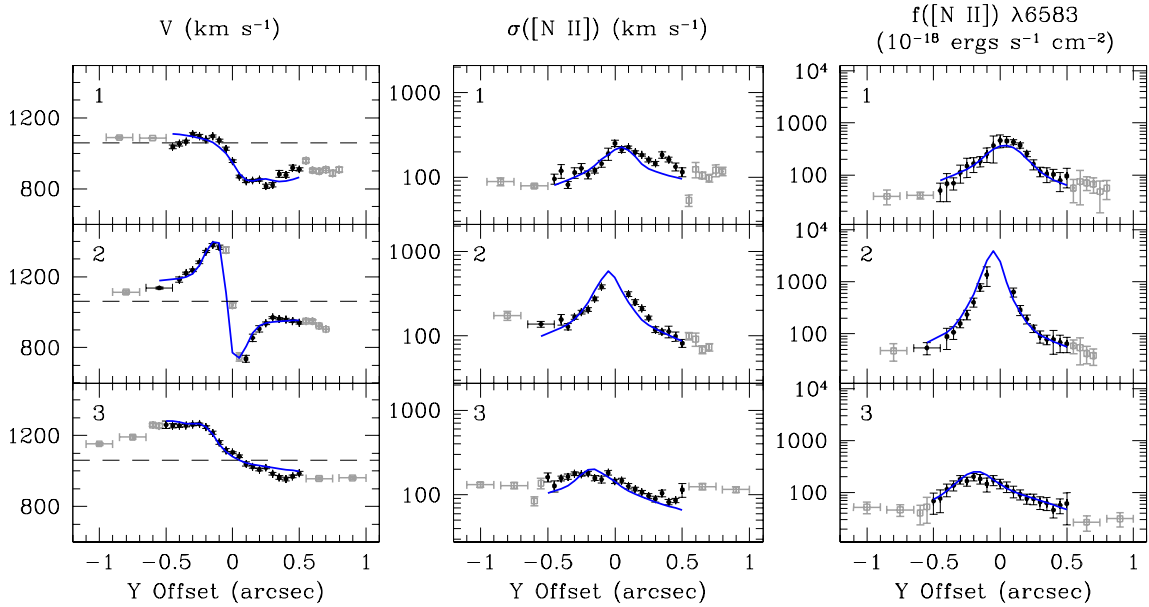}
\caption{Results of the final model with an asymmetric drift
correction; see Figure \ref{fig:bestfitmodel_noasymmdrift} for
description. \label{fig:bestfitmodel_asymmdrift}}
\end{center}
\end{figure*}


\begin{figure*}
\begin{center}
\plotone{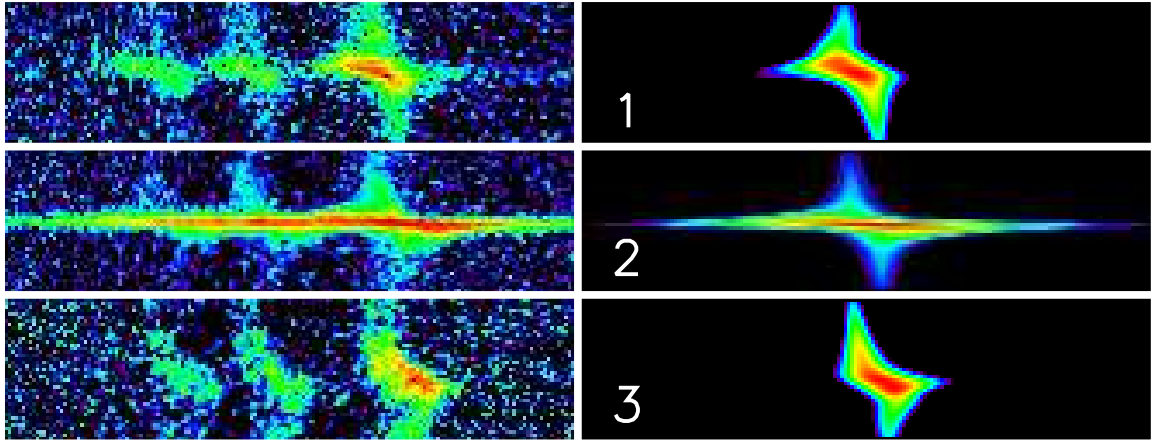}
\caption{Continuum-subtracted observed 2D STIS spectrum for the
H$\alpha$ + [\ion{N}{2}] region (left) and the synthetic 2D spectrum
of the [\ion{N}{2}] $\lambda 6583$ \AA\ emission line from the final
disk model with an asymmetric drift correction (right) for each of the
three slit positions. The spatial axis is vertical and the wavelength
axis is horizontal, with increasing wavelengths to the right. Each box
is 93 \AA\ in the dispersion direction and 2\farcs2 in the spatial
direction. The numbers correspond to the slit locations depicted in
Figure \ref{fig:wfpc2image}. \label{fig:2Dmodel}}
\end{center}
\end{figure*}

We explored the possibility that the intrinsic velocity dispersion
contributes dynamical support to the gas disk by including an
asymmetric drift correction in the model. Similar to the final model
without an asymmetric drift correction, we calculated the disk model
using a 0\farcs3-diameter PSF, $s = 10$, and $r_\mathrm{fit} = $
0\farcs5. We fixed $\Upsilon$ to 4 ($V$-band solar units), and
excluded the three uncertain central velocity measurements from the
fit.  We also included an intrinsic radial velocity dispersion profile
$\sigma_r$ as described in \S \ref{subsec:asymmdrift}.

The asymmetric drift correction depends on the radial gradient in the
number density of clouds in the disk, $\nu(r)$. We took $\nu$ to be
given by the intrinsic emission-line flux distribution, as has done in
the past \citep[e.g.,][]{Barth_2001, Coccato_2006, Neumayer_2007}.  In
order to compute an azimuthally-averaged value for the radial
derivative of $\nu$ in the asymmetric drift correction, we fit a
revised model to the emission-line surface brightness profile,
modeling it as an intrinsically circularly symmetric, inclined disk.
The model for $\nu(r)$ contained three Gaussians and one exponential
component (similar to that described in Table \ref{tab:sbparms}) but
with the modification that all four components were forced to have the
same centroid, ellipticity, and position angle.  Using this model for
$\nu(r)$, the asymmetric drift correction $v_c^2 - v_\phi^2$ was then
computed as a function of radius in each model calculation.  The disk
velocity field is then given by $v_\phi$.

When an asymmetric drift correction was included in the disk model
calculations, the black hole mass increased to $8.5 \times 10^8\
M_\odot$. This model is a better fit to the observed velocities, with
$\chi^2 = 524.8$ and $\chi^2_\nu = 10.1$, where 58 velocity
measurements were fit with 6 free parameters. In Figure
\ref{fig:bestfitmodel_asymmdrift}, we present the velocity, velocity
dispersion, and flux from the final model with an asymmetric drift
correction along with the observed velocity field. We also show the
observed 2D STIS spectrum and the synthetic spectrum from the final
model with an asymmetric drift correction for all three slit positions
in Figure \ref{fig:2Dmodel}. This figure demonstrates that the
rotating disk model is able to qualitatively match the complicated
nature of the observed spectra. In the right column of Table
\ref{tab:bestfitmodels} we provide the best-fit values of the model
parameters.

\subsection{Error Budget}
\label{subsec:errors}

We incorporated a number of sources of uncertainty in order to
determine the final range of possible black hole masses. In addition
to the formal model fitting uncertainties, we included the
uncertainties associated with the stellar mass-to-light ratio and
density profile, the PSF size, the subsampling factor, the fitting
region, the analytical form of the emission-line flux profile, and the
velocity measurements from the three central STIS rows. We
additionally explored the variation in the black hole mass when
several velocity measurements that appear to deviate from circular
rotation are excluded from the fit. For the disk model with an
asymmetric drift correction, we also included the contribution to the
uncertainty resulting from the parametrization of the $\nu$. These
sources of uncertainty are summarized below.

\emph{Model Fitting Uncertainty}: We began by estimating the model
fitting uncertainty. We explored the black hole mass parameter space
between $3.2 \times 10^8\ M_\odot$ and $6.2 \times 10^8\ M_\odot$ for
the disk model without an asymmetric drift correction, and from $(7.0
- 9.8) \times 10^8\ M_\odot$ for the model with the correction. We
held $M_\mathrm{BH}$ fixed while the remaining parameters ($i$,
$v_\mathrm{sys}$, $\theta$, $x_\mathrm{offset}$, and
$y_\mathrm{offset}$) were allowed to vary. Before calculating these
disk models, we rescaled the observed velocity uncertainties, which
has been the common practice in previous work (e.g.,
\citealt{Sarzi_2001, Barth_2001, deFrancesco_2006}, 2008). We added 29
km s$^{-1}$ in quadrature to each of the observed velocity
uncertainties in order to obtain $\chi^2_\nu \approx 1$ for the final
model without an asymmetric drift correction. Likewise, we acquired
$\chi^2_\nu \approx 1$ for the final model with an asymmetric drift
correction by adding 28 km s$^{-1}$ in quadrature to the observed
velocity uncertainties. By rescaling in this manner, the final
uncertainty on the black hole mass will increase. This approach
provides a conservative way in which to account for the detailed,
small-scale velocity structure that our disk models fail to reproduce.

\begin{figure*}
\begin{center}
\epsscale{0.8}
\plotone{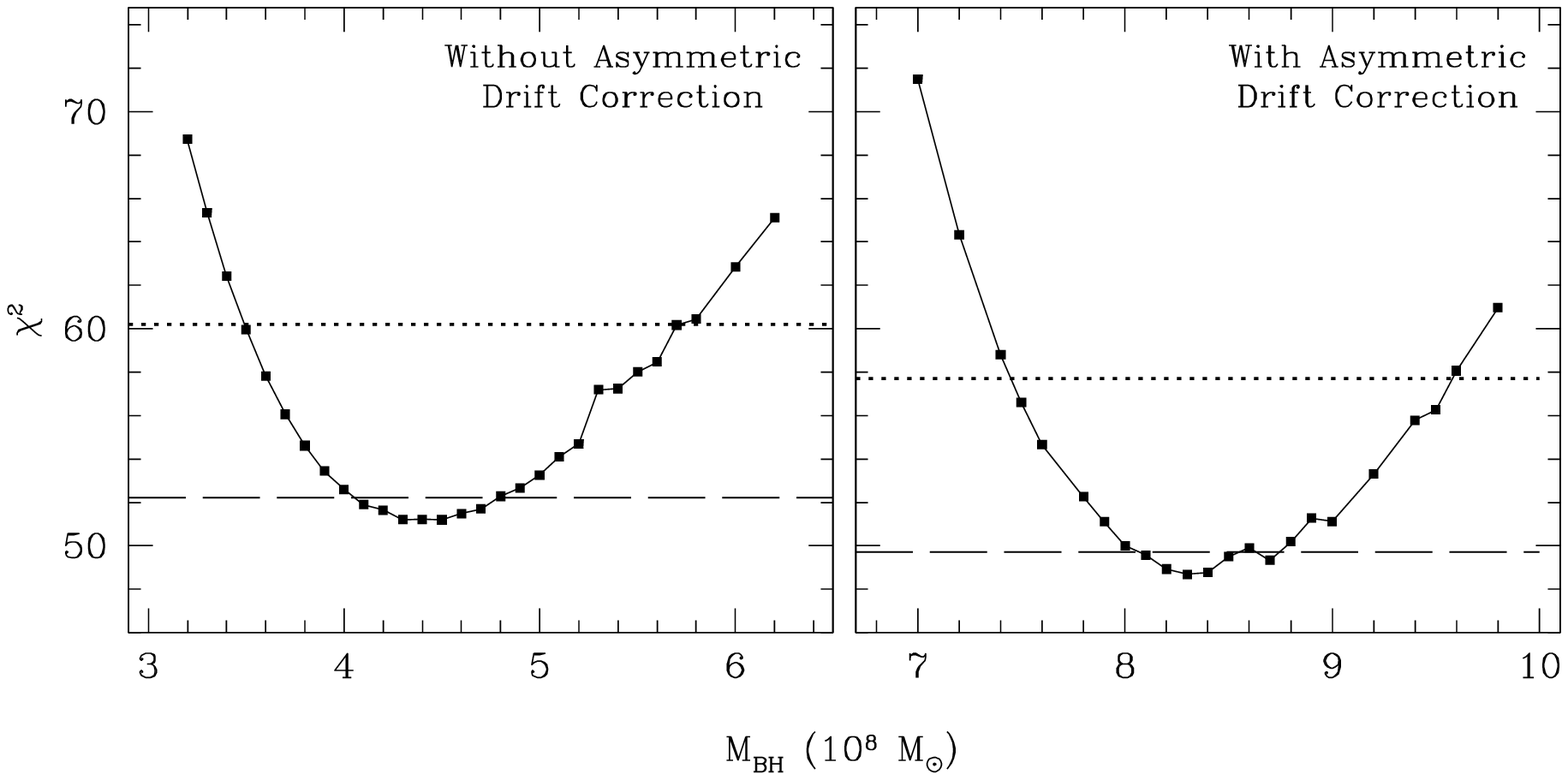}
\caption{Results of models without an asymmetric drift correction
(left) and including the correction (right) using a 0\farcs3-diameter
PSF, $s=10$, and $r_\mathrm{fit}=$ 0\farcs5. For each model,
$M_\mathrm{BH}$ was held fixed while $i$, $\theta$, $v_\mathrm{sys}$,
$x_\mathrm{offset}$, and $y_\mathrm{offset}$ were allowed to vary such
that $\chi^2$ was minimized. The dashed and dotted horizontal lines
denote where $\chi^2$ has increased by 1.0, and 9.0, from the minimum
value, corresponding to 1$\sigma$, and 3$\sigma$, uncertainties on
$M_\mathrm{BH}$, respectively. \label{fig:mbherr}}
\end{center}
\end{figure*}

Figure \ref{fig:mbherr} displays the results of the disk models. The
range of black hole masses which caused $\chi^2$ to increase by 1.0
from the minimum value provides the 68.3\% confidence limits
(1$\sigma$ uncertainties) on $M_\mathrm{BH}$. For the disk models
without an asymmetric drift correction, the 1$\sigma$ uncertainty
corresponds to $(4.0 - 4.8) \times 10^8\ M_\odot$, while the range of
black hole masses is $(8.0 - 8.8) \times 10^8\ M_\odot$ for the disk
models with the correction. The 99.7\% confidence limits (3$\sigma$
uncertainties) on $M_\mathrm{BH}$ were found by searching for the
range of black hole masses that caused the minimum $\chi^2$ to
increase by 9.0. The 3$\sigma$ uncertainties corresponded to $(3.5 -
5.7) \times 10^8\ M_\odot$ for the disk models without an asymmetric
drift correction and to $(7.4 - 9.6) \times 10^8\ M_\odot$ for the
models with the correction. These estimates also account for the small
uncertainties associated with the parameters $i$, $v_\mathrm{sys}$,
$\theta$, $x_\mathrm{offset}$, and $y_\mathrm{offset}$.


We found that rescaling the uncertainties on the observed velocities
resulted in a slightly different best-fit black hole mass than the
mass found from our final model with an asymmetric drift correction
presented in \S \ref{subsec:results_asymmdrift}. However, the black
hole mass found from our final model with an asymmetric drift
correction still falls within the 1$\sigma$ model fitting
uncertainty. Similar behavior has been seen in the previous analysis
of NGC 1300 and NGC 2748 by \cite{Atkinson_2005}. We choose to apply
the fractional model fitting error found above to the best-fit
$M_\mathrm{BH}$ from the final model presented in \S
\ref{subsec:results_asymmdrift}. Below, we describe the additional
contributions to the error budget.

\emph{Stellar Mass-to-Light Ratio and Density Profile}: Fixing
$\Upsilon = 4$ does not have a large effect on the black hole mass,
further emphasizing that the STIS data is fairly insensitive to the
stellar contribution. We find that by allowing $\Upsilon$ to vary
during the fit, the best-fit stellar mass-to-light ratio is $\Upsilon
= 0$ and the black hole mass is $M_\mathrm{BH} = 4.8 \times 10^8\
M_\odot$ for the model without an asymmetric drift correction. Thus,
increasing the mass-to-light ratio from the best-fit value of
$\Upsilon = 0$ to $\Upsilon = 4$ results in a 12\% decrease in
$M_\mathrm{BH}$. For the model with an asymmetric drift correction,
allowing $\Upsilon$ to vary during the fit caused $M_\mathrm{BH}$ to
change by only 1\% from the best-fit mass, while $\Upsilon$ decreased
to 2.6 $V$-band solar units.

Also, as discussed in \S \ref{subsec:stars} and shown in Figure
\ref{fig:stellarvc}, the uncertainty related to the exclusion of the
central compact component from the mass budget is unlikely to affect
our $M_{\rm BH}$ measurement. In fact, even if all the nuclear light
is assigned a mass meaning, the black hole mass remains the same at
$8.5 \times 10^8\ M_{\odot}$ and drops to $4.1 \times 10^8\ M_{\odot}$
(a 5\% change from the best-fit mass), with and without asymmetric
drift correction, respectively.

\emph{PSF size}: When PSFs of different sizes (0\farcs3 -- 0\farcs6 in
diameter) were used in the calculation of the model without an
asymmetric drift correction, the black hole mass varied between $(4.3
- 4.5) \times 10^8\ M_\odot$ with a root-mean-square (rms) scatter of
$8.7 \times 10^6\ M_\odot$. This corresponds to 2\% of the best-fit
black hole mass. Similarly, for the model with an asymmetric drift
correction, the black hole mass varied between $8.5 \times 10^8\
M_\odot$ and $8.6 \times 10^8\ M_\odot$ with an rms scatter equal to
$4.3 \times 10^6\ M_\odot$, or 1\% of the best-fit black hole mass.

\emph{Subsampling Factor}: For disk models without an asymmetric drift
correction, using subsampling factors ranging from $s=2$ to $s=10$
resulted in black hole masses between $M_\mathrm{BH} = 4.1 \times
10^8\ M_\odot$ and $4.3 \times 10^8\ M_\odot$ with a rms scatter of
only $7.5 \times 10^6\ M_\odot$, or 2\% of the best-fit black hole
mass. The black hole mass ranged from $(8.2 - 8.6) \times 10^8\
M_\odot$ for disk models with an asymmetric drift correction. The rms
scatter was $1.4 \times 10^7\ M_\odot$, which is 2\% of the best-fit
black hole mass.

\emph{Fitting Region}: For the disk models without an asymmetric drift
correction, the black hole mass varied between $M_\mathrm{BH} = 4.2
\times 10^8\ M_\odot$ and $4.5 \times 10^8\ M_\odot$ for fitting
regions between 0\farcs3 and 1\arcsec. The rms scatter in
$M_\mathrm{BH}$ was small at $1.3 \times 10^7\ M_\odot$, or 3\% of the
best-fit black hole mass. The best-fit black hole mass for models with
an asymmetric drift correction varied between $(8.2 - 8.7) \times
10^8\ M_\odot$, with an rms scatter that amounted to $1.6 \times 10^7\
M_\odot$, corresponding to 2\% of the best-fit mass.

\emph{Emission-Line Flux}: The dynamically cold, thin-disk models with
different parametrizations of the emission-line flux (previously
described in \S \ref{sec:modelresults}) returned best-fit masses
between $4.2 \times 10^8\ M_\odot$ and $4.7 \times 10^8\ M_\odot$. The
rms scatter was $1.6 \times 10^7\ M_\odot$, or 4\% of the best-fit
black hole mass. When the different emission-line flux models were
used to weight the line-of-sight velocity profiles in the model with
an asymmetric drift correction, we found that the black hole mass
fluctuated between $8.1 \times 10^8\ M_\odot$ and $8.9 \times 10^8\
M_\odot$ with an rms scatter of $2.7 \times 10^7\ M_\odot$, or 3\% of
the best-fit $M_\mathrm{BH}$.

\emph{Velocity Measurements Within 0\farcs05 of the Nucleus}: We
calculated a disk model without an asymmetric drift correction in
which the three innermost velocity measurements were included in the
fit, as a comparison to the final model that excludes these uncertain
measurements. The black hole mass decreased by just 2\% to
$M_\mathrm{BH} = 4.2 \times 10^8\ M_\odot$. The inclusion of the
velocity measurements from the three innermost STIS rows also had an
insignificant impact on $M_\mathrm{BH}$ for the model with asymmetric
drift correction; $M_\mathrm{BH}$ decreased by 1\% from the best-fit
mass. Because the sphere of influence is well resolved by the STIS
data, excluding the three central velocity measurements during the fit
has a small effect on the black hole mass.

\emph{Excluding Several Velocity Measurements that Deviate from
Circular Rotation:} The final models with and without an asymmetric
drift correction match the overall shape of the observed velocity
curves well, but are unable to reproduce all of the velocity features,
resulting in moderate $\chi^2_\nu$ values of 10.1 and 11.4,
respectively. In order to measure the impact of several velocity
measurements that show localized departures from pure circular
rotation on the black hole mass and $\chi^2_\nu$, we fit models with
and without an asymmetric drift correction to various data sets. We
constructed data sets in which we removed anywhere from a single
velocity measurement to ten velocity data points that appeared to be
discrepant from the final models presented in \S
\ref{subsec:results_noasymmdrift} and \S
\ref{subsec:results_asymmdrift}, such as the points at Y-Offset -0.45"
to -0.35" from slit position 1, the point at Y-Offset -0.55" from slit
position 2, and the points at Y-Offset 0.30" to 0.45" from slit
position 3. For the model without an asymmetric drift correction, we
found that the black hole mass varied between $4.1 \times 10^8\
M_\odot$ and $4.3\times 10^8\ M_\odot$, with an rms deviation of just
2\% of the best-fit mass, and the lowest $\chi^2_\nu$ was
6.1. Similarly, for the model with an asymmetric drift correction, the
rms deviation in the black hole mass was 1\% of the best-fit mass,
varying between $(8.3 - 8.5) \times 10^8\ M_\odot$, with the lowest
$\chi^2_\nu$ being 5.3. Thus, the quality of the fit to the observed
velocity curves improves significantly, while the black hole mass
remains roughly constant. This lends further support that the best-fit
masses from our final models with and without an asymmetric drift
correction are trustworthy despite the moderate $\chi^2_\nu$ values.

\emph{Number Density of Clouds}: For models with an asymmetric drift
correction, we also tested a variety analytical functions for
$\nu$. These distributions included both shallow and steep profiles,
and ranged in complexity with anywhere from two to six components. The
functions were composed of intrinsic circularly-symmetric Gaussians
and exponentials. When calculating the models, we continued to weight
the line-of-sight velocity profiles by the 3 Gaussians $+$ 1
exponential emission-line flux model (model A) discussed in \S
\ref{subsec:emissionline_sb}. The effect on the black hole mass was
small, and we measured masses between $(7.1 - 8.6) \times 10^8\
M_\odot$ with an rms scatter of $5.6 \times 10^7\ M_\odot$, which is
7\% of the best-fit black hole mass.

All of the above sources of uncertainty were added in quadrature to
the model-fitting uncertainty. The final range of black hole masses
for the cold thin-disk model is $(3.6 - 5.1) \times 10^8\ M_\odot$
(1$\sigma$ uncertainties) and $(3.3 - 5.8) \times 10^8\ M_\odot$
(3$\sigma$ uncertainties) with a best-fit black hole mass of $4.3
\times 10^8\ M_\odot$. For the disk model with an asymmetric drift
correction, the final range of masses is $(7.7 - 9.4) \times 10^8\
M_\odot$ (1$\sigma$ uncertainties) and $(7.3 - 10.0) \times 10^8\
M_\odot$ (3$\sigma$ uncertainties) with a best-fit black hole mass of
$8.5 \times 10^8\ M_\odot$.

\section{Discussion}
\label{sec:discussion}

By modeling the M84 emission-line gas kinematics as a dynamically
cold, thin disk in circular rotation, we find a best-fit black hole
mass of $(4.3^{+0.8}_{-0.7}) \times 10^8\ M_\odot$. Incorporating an
asymmetric drift correction in our disk model results in a best-fit
black hole mass of $(8.5^{+0.9}_{-0.8}) \times 10^8\ M_\odot$. We
favor the disk model with an asymmetric drift correction because it is
physically plausible that the intrinsic turbulence affects the disk's
dynamics and also because the model provides a better fit to the
observed radial velocities.

\begin{figure}
\begin{center}
\epsscale{1.1}
\plotone{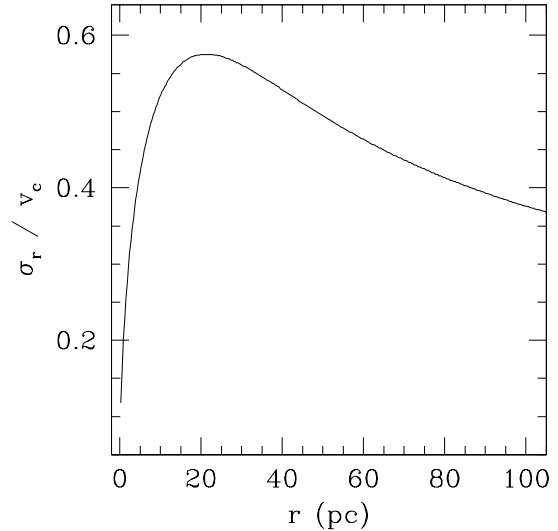}
\caption{Ratio of $\sigma_r$ to $v_c$ as a function of distance from
the M84 nucleus for the final model discussed in \S
\ref{subsec:results_asymmdrift}. \label{fig:sigmarvc}}
\end{center}
\end{figure}

There are a few caveats associated with the asymmetric drift
correction worth mentioning. Our treatment of the asymmetric drift
correction is only an approximation, and the black hole mass
measurement becomes increasingly less certain as $\sigma_r/v_c$
approaches $\sim 1$. In the case of M84, we find that $\sigma_r/v_c$
can reach moderate values of 0.57 (see Figure
\ref{fig:sigmarvc}). Moreover, the increase in the black hole mass
when an asymmetric drift correction is applied is extremely large (a
98\% change), and can be attributed to the very steep radial gradients
in both $\nu$ and $\sigma_r$. While $\sigma_r$ is known fairly well
through a fit to the observed line widths, the radial distribution of
the number density of clouds in the disk is not known. Following past
work, we have used the emission-line flux profile as a proxy for the
density profile $\nu(r)$. While this is at best a rough approximation,
we found that different parametrizations of $\nu$ did not result in
substantial changes to the inferred black hole mass. However, if
instead a very simplistic asymmetric drift correction is applied,
where $v_c^2 - v_\mathrm{rot}^2 = \sigma_r^2$, the black hole mass
decreases by $\sim 30$\% to $M_\mathrm{BH} = 6.0 \times 10^8\
M_\odot$. Finally, the asymmetric drift correction is strictly
applicable in the limit of collisionless particles, as in stellar
dynamics. Here, we are applying the correction to gas clouds, which
are not collisionless particles, and therefore the analogy is not
perfect, although it is still a useful approximation
\citep[e.g.,][]{Valenzuela_2007}. Therefore, calculating a model which
includes an asymmetric drift correction is an informative exercise,
but only provides an approximate indication of the dynamical influence
of the intrinsic velocity dispersion.

Additionally, dust obscuration is an issue in M84 and affects both our
gas dynamical models. As can be seen in the WFPC2 $V$-band image in
Figure \ref{fig:wfpc2image}, there are two prominent dust lanes within
the central $\sim 4$\arcsec\ that are oriented roughly east to
west. Based on a $V - I$ color map, \cite{Bower_1997} measure a dust
mass of $7 \times 10^4\ M_\odot$, and note that the dust extinction is
the greatest along the northern portion of the central dust lane and
at a patch located on the nucleus. The effects of dust on the
determination of the stellar mass profile should be minimized, as the
\cite{Kormendy_2009} surface brightness profile uses NICMOS $H$-band
data for the nuclear region. Also, as we have shown above, the model
fits are insensitive to the stellar mass contribution because the
black hole dominates over the region we are modeling. Dust extinction
will affect the observed gas kinematics, however including these
effects in the calculation requires radiative transfer models of the
disk's gas dynamics. Such modeling has yet to be applied to gas
kinematical measurements and is beyond the extent of this paper.


The best-fit black hole mass of $(8.5^{+0.9}_{-0.8}) \times 10^8\
M_\odot$ is a factor of $\sim 2$ smaller than the mass measured by
\cite{Bower_1998} of $(1.5^{+1.1}_{-0.6}) \times 10^9\ M_\odot$,
although the two measurements are formally consistent within their
quoted uncertainties.  If we consider our best-fitting model that does
not include asymmetric drift (perhaps a more fair comparison since
Bower et al.\ did not include asymmetric drift), our result is a
factor of $\sim3$ smaller than theirs.  While there are many factors
that contribute to this difference, one reason is likely to be the
different assumptions we have made about the structure of the velocity
field.  Bower et al.\ decomposed the emission lines into two distinct
kinematic components, which they characterized as one rotating disk
component and one unrelated, low-velocity component, while we have
modeled the kinematics as arising from a single rotating disk.  The
observed and modeled line profiles of the central slit position, shown
in Figure \ref{fig:2Dmodel}, help to illustrate the situation.  Close
to the nucleus, the emission lines flare out into a broad ``fan'' of
emission with large line widths and complex, strongly non-Gaussian
profiles, due to a combination of instrumental and rotational
broadening, and the increase in the intrinsic velocity dispersion
close to the nucleus.  \cite{Maciejewski_Binney_2001} describe in
detail how this fan of emission, which can mimic the appearance of two
separate kinematic components, can be generated through the
combination of rotational broadening and the effects of observing
through a slit that is wider than the PSF core.

Moreover, our model calculation includes a more detailed treatment of
the telescope and spectrograph optics than was carried out by
\citet{Bower_1998}, and includes the full propagation of emission-line
profiles through the spectrograph, rather than a simpler propagation
of mean velocities.  While it is difficult to pinpoint the exact cause
of the difference in the resulting black hole mass, these are likely
to be the primary reasons.  Our results show that we can obtain a
robust fit of a disk model to the data even in the absence of any
measurements from the central three rows, where the line profiles are
most severely blended and asymmetric and the velocity field is most
strongly affected by PSF blurring.

Of the remaining disk model parameters, our best-fit systemic velocity
and inclination angle of $v_\mathrm{sys} = 1060$ km s$^{-1}$ and $i =
72^\circ$ differ from the results of \cite{Bower_1998}, who found
$v_\mathrm{sys} = 1125$ km s$^{-1}$ and $i = 80^\circ$. The systemic
velocity determined through our modeling is in better agreement with
the recession velocities given by the NASA Extragalactic Database
(NED). The velocities quoted by NED were measured from optical lines,
and ranged in value from 954 -- 1119 km s$^{-1}$, with an average of
1025 km s$^{-1}$. \cite{Bower_1998} did not fit for the relative angle
between the STIS slit and the major axis of the gas disk. Rather, they
took the disk position angle to be that of the major axis of the
large-scale emission-line structure (83$^\circ$ east of north)
measured by \cite{Baum_1998}. In our disk models, we allowed $\theta$
to be a free parameter and found a similar position angle for the
major axis of the gaseous disk. Our best-fit value is $\theta =
28^\circ$, corresponding a disk major axis position angle of
76$^\circ$ east of north. Also, \cite{Bower_1998} adopted a stellar
mass-to-light ratio of $\Upsilon = 5$ ($V$-band solar units),
comparable to the one fixed in our model of $\Upsilon = 4$ ($V$-band
solar units).

The comparison of our results with those of
\cite{Maciejewski_Binney_2001} is less direct, since they did not
carry out model fitting to the observed velocity field.  Instead, they
estimated the black hole mass based on the visual location of a
caustic feature in the spectrum of the central slit position.  This
caustic results from the interplay between the disk rotation and the
gradient in instrumental wavelength shifts across the slit.  These two
line-broadening effects can be oppositely directed and can effectively
cancel each other out, resulting in a sharp reduction in the observed
line width at a specific location in the disk.  We do not clearly see
evidence for this caustic feature in the 2D spectrum or in the
measured emission-line widths, so we cannot reproduce their estimate
of $M_\mathrm{BH}$.  While such caustics should in principle occur in
STIS spectra of emission-line disks, it is possible that the intrinsic
line width in the M84 disk is so large that it washes out any
observable signature of the caustic.  We note, however, that our
best-fitting model without asymmetric drift gives a result that is
consistent with that of \cite{Maciejewski_Binney_2001}.

With respect to the $M_\mathrm{BH} - \sigma_\star$ and $M_\mathrm{BH}
- L$ relationships, our best-fit black hole mass for M84 of
$(8.5^{+0.9}_{-0.8}) \times 10^8\ M_\odot$ is a factor of $\sim 2$
smaller than the mass from \cite{Bower_1998} that has been most often
adopted in the relationships. This new mass measurement lies closer to
the mass expected from the $M_\mathrm{BH} - \sigma_\star$ and
$M_\mathrm{BH} - L$ relationships measured by
\cite{Gultekin_2009}. Recent stellar dynamical studies of other
galaxies, as well as our analysis of M84, show that some previous
black hole mass measurements should be re-evaluated, and determination
of the true shape and scatter of the $M_\mathrm{BH} - \sigma_\star$
and $M_\mathrm{BH} - L$ relationships is an evolving process that will
continue to change as observations and modeling techniques continue to
improve.  Recent calibrations of the relations still include some of
the early gas-dynamical measurements that were done with the FOS on
\emph{HST}, and these should be particularly important targets to
revisit with future observations.  Equally important are cross-checks
between gas-dynamical and stellar-dynamical techniques within the same
object, as each mass measurement method suffers from independent
systematic uncertainties. In particular, gas-dynamical models would
benefit tremendously from a better understanding of the physical
origin and the dynamical influence of the intrinsic velocity
dispersion.

\section{Conclusions}
\label{sec:conclusions}

With the goal of resolving the uncertainty in the M84 black hole mass,
we have re-analyzed multi-slit archival \emph{HST} STIS observations
of the nuclear region of M84. We mapped out the 2-dimensional
[\ion{N}{2}] $\lambda 6583$ emission-line velocity, velocity
dispersion, and flux. We then modeled the velocity field as a cold,
thin disk in circular rotation. The line widths predicted by this
model are smaller than the observed line widths, and we found that an
intrinsic velocity dispersion is needed in order to match the
observations. The additional velocity dispersion may have a dynamical
origin and provide pressure support to the gaseous disk, thus we
calculated a second disk model with an asymmetric drift correction. We
found that the disk model with an asymmetric drift correction is a
better fit to the data than the cold thin-disk model. This model gives
a black hole mass of $(8.5^{+0.9}_{-0.8}) \times 10^8\ M_\odot$
(1$\sigma$ uncertainties).

We have employed a more rigorous and comprehensive gas-dynamical model
to measure the black hole mass in M84 than the previous two
studies. In addition to calculating an asymmetric drift correction, we
included a more sophisticated treatment of the effects of the
telescope optics and interpreted the complex nuclear spectra as
arising from a single gas component. We also performed an error
analysis that encompasses a number of sources of uncertainty in order
to determine the range of possible black hole masses.

Our new mass measurement for M84 is a factor of $\sim 2$ smaller than
the \cite{Bower_1998} value of $M_\mathrm{BH} = (1.5^{+1.1}_{-0.6})
\times 10^9\ M_\odot$ and a factor of $\sim 2$ larger than the
\cite{Maciejewski_Binney_2001} estimate of $M_\mathrm{BH} = 4 \times
10^8\ M_\odot$. The black hole mass given by \cite{Bower_1998} is
often used in the $M_\mathrm{BH} - \sigma_\star$ and $M_\mathrm{BH} -
L$ relationships. Therefore, while recent stellar-dynamical work has
found that several masses at the high-end of the black hole mass-host
galaxy relationships have been underestimated by a factor of $\sim 2$,
we find that the gas-dynamical black hole mass measurement for M84 has
been overestimated by a factor of $\sim 2$. With this adjustment to
the black hole mass, M84 lies closer to the mass expected from both
the $M_\mathrm{BH}-\sigma_\star$ and $M_\mathrm{BH}-L$
relationships. Future work should thus aim to understand the
systematics associated with each of the main mass measurement methods,
and to re-examine those objects for which uncertain mass measurements
remain.

\acknowledgements

Research by A.~J.~B. and J.~L.~W. has been supported by NSF grant
AST-0548198.  We thank Luis Ho for stimulating discussions that helped
to motivate this project.  This research has made use of the NASA/IPAC
Extragalactic Database (NED) which is operated by the Jet Propulsion
Laboratory, California Institute of Technology, under contract with
NASA.  The data presented in this paper were obtained from the
Multimission Archive at the Space Telescope Science Institute
(MAST). STScI is operated by the Association of Universities for
Research in Astronomy, Inc., under NASA contract NAS5-26555.

\end{document}